\documentclass[a4paper, 12pt]{article}

\usepackage{amsmath,amssymb,amsthm}
\usepackage{verbatim}
\usepackage{amsfonts}
\usepackage{mathtools}
\usepackage{graphicx}
\usepackage{tikz}
\usepackage{threeparttable}
\usepackage{empheq}
\usepackage{lmodern}
\usepackage{booktabs}
\usepackage{multirow}
\usepackage{tabularx}
\usepackage{natbib}
\usepackage{pifont}
\usepackage{geometry}
\usepackage{authblk}
\usepackage{mwe}
\usepackage{longtable}
\usepackage{systeme}
\usepackage{mathrsfs}
\usepackage{enumitem}
\usepackage{csquotes}
\usepackage{titlesec} 
\usepackage{etoolbox}
\usepackage{soul} 
\usepackage{float}
\usepackage{setspace}
\usepackage{algorithm}
\usepackage{neuralnetwork}
\usepackage{algpseudocode}
\usepackage{booktabs}
\usepackage{multirow}
\usepackage[UKenglish]{babel}
\usepackage[UKenglish]{isodate}
\usepackage[section]{placeins}
\usepackage[font=small,labelfont=bf]{caption}
\usepackage{rotating} 
\usepackage{hyperref} 
\usepackage{caption}
\usepackage{subcaption}

\usepackage{todonotes}

\hypersetup{colorlinks=true,citecolor=blue, urlcolor=blue}

\newlist{level}{itemize}{4}

\setlist[level]{label={},noitemsep,topsep=0pt}

\newcommand\blfootnote[1]{%
  \begingroup
  \renewcommand\thefootnote{}\footnote{#1}%
  \addtocounter{footnote}{-1}%
  \endgroup
}

\newcommand\iidN{\overset{iid}{\sim}\mathcal{N}} 

\DeclareMathOperator*{\argmax}{argmax}

\begin{document}


\title{Deep Dynamic Factor Models}

\author[1]{Paolo Andreini}
\author[1]{Cosimo Izzo}
\author[2]{Giovanni Ricco}
\affil[1]{\it\small Independent Researcher}
\affil[2]{\it\small CREST \`Ecole Polytechnique, University of Warwick, CEPR, OFCE-SciencesPo}

\date{\vspace{0.5cm}\small{First Version: February 2020\\This version: \today}}
\cleanlookdateon


\maketitle
\blfootnote{
	This work reflects the analysis and views of the authors, Paolo Andreini, Cosimo Izzo and Giovanni Ricco. No reader should interpret this work to present the views of any third party. Assumptions, opinions, views and estimates constitute the authors' judgment as of the date given and are subject to change without notice and without duty to update. 
	
	The replication code for the simulations of this paper is available on the \href{https://github.com/cosimoizzo/DDFM}{GitHub repository}.
	
		We are grateful to Matteo Barigozzi, Antonello D'Agostino, Aldo Lipani, Massimiliano Marcellino, Jasper McMahon, Francesca Medda, Ramin Okhrati, Filippo Pellegrino, Ivan Petrella, Giuseppe Ragusa and Lucrezia Reichlin and to the conference participants of the the 2nd Vienna Workshop on Economic Forecasting 2020, the 40th International Symposium on Forecasting, the 18th Real-Time Data Analysis Methods and Applications conference, and the CFE-CMStatistics 2022 conference for many helpful suggestions and comments. }
\thispagestyle{empty}

\vspace{-0.5cm}
\begin{abstract}
A novel deep neural network framework -- that we refer to as Deep Dynamic Factor Model (D$^2$FM) --, is able to encode the information available, from hundreds of macroeconomic and financial time-series into a handful of unobserved latent states. While similar in spirit to traditional dynamic factor models (DFMs), differently from those, this new class of models allows for nonlinearities between factors and observables due to the autoencoder neural network structure. However, by design, the latent states of the model can still be interpreted as in a standard factor model. Both in a fully real-time out-of-sample nowcasting and forecasting exercise with US data and in a Monte Carlo experiment, the D$^2$FM improves over the performances of a state-of-the-art DFM.
\end{abstract}
\vspace{0.2cm}




\newpage
\renewcommand{\baselinestretch}{1.5}
\normalsize

\section{Introduction} 

An overarching idea in macroeconomics, already shaping the work of \cite{RePEc:nbr:nberbk:burn46-1}, is that a few common forces can explain the joint dynamics of many macroeconomics variables. This stylised view of the economic data generating process has long informed the effort of economic modelling -- for example, in the Real Business Cycle (RBC) and Dynamic Stochastic General Equilibrium (DSGE) literature -- and is one of the very few robust facts in empirical macroeconomics, motivating the use of factor models \citep[see for example][]{STOCK2016415}.

In macroeconometrics, factor models were firstly introduced by \cite{geweke1977dfm} and \cite{sargentsims1977dfm} and are a very early instance of `big data' in macroeconomics. Dynamic Factor Models (DFMs) deal with a large cross-section of data (`large N problem') by applying a linear dynamic latent state framework to the analysis of economic time-series. The underlying assumption of these models is that there is a small number of pervasive unobserved common factors that stir the economy and inform the comovements of hundreds of economic variables. Economic times series are also possibly affected by variable-specific (idiosyncratic) disturbances. These idiosyncratic disturbances can be due to either measurement error or variable-specific disturbances. Dynamic factor models are workhorse models in macroeconometrics and a large body of empirical evidence has found that, in many applications, a small number of factors -- as many as two -- can capture a dominant share of the variance of all the key macroeconomic and financial variables.\footnote{This family of models has been applied intensively in econometrics to different problems such as forecasting, structural analysis and the construction of economics activity indicators (see, among many others, \citealp{stockwatson2002a, stockwatson2002b, forni2001generalized, forni2000generalized, forni2005generalized, forni2015dynamic, forni2018dynamic, altissimo2010new}).} 

Factor models are robust and flexible models, also able to accommodate for missing observations, jagged patterns of data and mixed frequencies.\footnote{Jagged edges arise when there is a varying number of missing observations at the end of multiple time-series due to non-synchronous release dates.} However, two of their important limitations are (i) the almost always assumed linear structure, and (ii) the limited scalability of these models due to the computational challenges that are encountered when estimating factors models with more than a few dozens of variables. 

This paper introduces a generalisation of factor models in a deep learning framework -- which we label Deep Dynamic Factor Models (D$^2$FMs) -- that deals effectively with these challenges, while maintaining the same degree of flexibility and of interpretability of a standard DFM. Indeed, our deep learning model can `encode'  the information about the state of the economy, as available in real-time, from hundreds of macroeconomic and financial variables at mixed frequency and with `jagged edges', into a handful of unobserved latent states. While similar in spirit to traditional dynamic factor models, differently from those, our model allows for non-linearities both in the encoding -- from variables to factors -- and in the decoding map -- back to the variables from the factors. We also discuss how to generalise it further to nonlinear factor dynamics.

The backbone of our modelling approach is provided by a dynamic autoencoder structure capturing the common information across variables, and whose parameters are estimated via gradient-based backpropagation. An autoencoder is a type of unsupervised learner that maps a number of variables (`input layer') into themselves (`output layer') by first `encoding' the variables' common information into a lower dimensional `code' (viz. non-linear factors), and then `decoding' it. Autoencoders are formed by a series of internal (hidden) layers each formed by a number of nodes (neurons). The encoding happens in the first half of the model, and it is the results of a series of non-linear transformations of linear combinations of inputs coming from the previous layer to the current one. Each neuron in each layer operates one of these transformations. The sequence of layers provides `depth' to the neural net, while the number of neurons per layer provides `width'. Autoencoders can be thought of as a nonlinear generalisation of principal component analysis \citep[see][]{hinton2006reducing}, which are able to transform very high-dimensional data into low-dimensional factors without having to assume a linear factor structure.\footnote{As discussed in \cite{baldi1989neural}, affine decoder and encoder without any nonlinearity and squared error loss will recover the same subspace of PCA. Moreover, when nonlinearity is added into the encoding network, PCA appears as one of the many possible representations \citep[see][]{bourlard1988auto, japkowicz2000nonlinear}.} 

The central methodological contribution of our paper is to show how to embed autoencoders in a dynamic nonlinear factor model structure with idiosyncratic components to tackle macroeconomic problems, thus providing generalisation to linear factor models. The equivalence between maximum likelihood estimation and minimisation of mean squared error together with the Universal Approximation Theorem (\citealp{cybenko1989approximation}, \citealp{HORNIK1989359,HORNIK1990551}) allow to reinterpret the D$^2$FMs and the procedure adopted in estimating them as an efficient computational method to approximate the maximum likelihood estimates of a general nonlinear factor models. A second contribution of this paper is to show how to incorporate in this framework general patterns of missing data, jagged edges and mixed frequencies, by extending gradient-based backpropagation methods for autoencoders.

Our methodology is computationally efficient and provides large gains in terms of computational speed when compared to maximum likelihood methods for DFMs. At the best of our knowledge, this paper is the first to adopt an autoencoder structure in a dynamic model with both factor dynamics and dynamic idiosyncratic components, in a state-space framework for real-time high dimensional mixed frequencies time-series data with arbitrary patterns of missing observations.

The proposed D$^2$FM  framework is very general and can be, in principle, applied to many different problems both in forecasting and in structural analysis, as done with DFMs.  Indeed, the model is designed to be in spirit as close as possible to DFMs.  

We test the performances of the D$^2$FM both in a controlled environment through Monte Carlo experiments, and empirically on US data in a forecasting/nowcasting exercise in the spirit of \cite{giannone2008nowcasting}. The Monte Carlo experiments show that the D$^2$FM largely outperforms the state-of-the-art DFM when the true data generating process is nonlinear, and offers similar performances in the linear case.

In the empirical application, we employ the D$^2$FM to encode a full real-time version of the \cite{mccracken2016fred}'s FRED-MD `big data' dataset for the US.  The specification of the model and its hyperparameters are not fixed ex-ante but are instead selected in an intensive cross-validation exercise. The model is then evaluated in real-time by comparing its forecasting, nowcasting and backcasting performances against two benchmarks: (i) a univariate AR(1) model; and  (ii) a state-of-the-art DFM estimated using a quasi maximum likelihood and an Expectation-Maximisation algorithm (see \citealp{giannone2008nowcasting} and \citealp{banbura2014maximum}). The D$^2$FM  outperforms the two benchmarks, in forecasting and in nowcasting, with gains of up to around a 14\% improvement when measured in terms of the root mean square forecast errors (RMSFE).\newline

The paper is organised as follows. The remainder of this section discusses the related literature. Section \ref{sec:background} provides some background and the core intuition guiding our methodology: the idea that autoencoders can be seen as static generalisations of PCA, and hence that dynamic versions of these models should be seen as nonlinear generalisations of linear dynamic factor models. Section \ref{sec:estimation_sec} presents the methodology proposed and discuss its estimation. Section \ref{sec:mc_sim} illustrates the Monte Carlo experiments. In section \ref{sec:proposed_model} we discuss how to deal with the specificities of economic data. Section \ref{sec:empirical_app} describes the empirical application which has the aim to track in real-time the US GDP, using real-time data. Section \ref{sec:conclusion} summarises the main results of the paper and sketches some possible future path of research. Additional technical details and a data description are provided in the Appendix.\newline

{\bf Related Literature} This work connects three distinct areas of research: the econometric literature focused on dynamic factor model estimation, research on non-linearity in empirical macro models, and deep learning research on autoencoders and latent factor models for analysing time series data.

The key problem in the factor model literature is that, due to the latency of the factors, maximum likelihood estimators cannot be derived explicitly. \cite{geweke1977dfm} and \cite{sargentsims1977dfm} proposed optimised algorithms for small models in the frequency domain, while \cite{engle1981one} and \cite{stock1989new} offered solutions in the time domain.\footnote{Specifically, \cite{engle1981one} estimate the dynamic factor model using a state-space representation in which they apply the Kalman filter to compute the likelihood used for the full maximum likelihood estimation of the parameters. \cite{watson1983alternative} and \cite{shumway1982approach} adapt the Expectation Maximisation (EM) algorithm of \cite{dempster1977maximum} for state-space representation allowing the presence of missing data, but only in the specific case where the matrix of the measurement equation is known.} The common drawback of all these proposed methods is that, in general, the maximum likelihood approach is unfeasible for datasets where the cross-section size is large. To solve this problem in \cite{forni1998let}, \cite{stockwatson2002a} and \cite{giannone2008nowcasting}  have proposed non-parametric methods based on the principal component analysis to estimate the latent components with large cross-section data.\footnote{\cite{doz2012quasi} and \cite{barigozzi2019quasi} have shown that when the size of the cross-section tends to infinite the estimates obtained by a quasi-maximum likelihood approach are consistent, also when there is a weak cross-sectional correlation in the idiosyncratic components.} The intuition driving this literature and its applications to macroeconomic problems inform our approach, which can be viewed as providing both a generalisation of standard dynamic factor models, and an efficient numerical approach for handling big data.

Our work also connects to the econometric literature that has explored the extent of nonlinearities in macroeconomic data and proposed univariate and multivariate nonlinear time-series models (\citealp[see][for a comprehensive literature review]{kock2011forecasting}). An important part of this literature has estimated dynamic latent models with nonlinearities, which are explicitly modelled through structural breaks, Markov switching regression or threshold regression (see \citealp{barnett2016real}, \citealp{camacho2012markov}, \citealp{marcellino2010factor}, \citealp{korobilis2006forecast} and \citealp{nakajima2013bayesian}).\footnote{Nonlinearities have been also modelled in structural factor models using DSGE models (Dynamic Stochastic General Equilibrium models) as in \cite{amisano2011exact} to detect regime switching in volatility.} In this literature, the approach of \cite{BAI2008304} is the closest in spirit to ours and an important early effort at including nonlinearities in factor models. In that paper, either principal components of nonlinear transformation of the data are estimated or nonlinear transformation of the factors are added to a linear factor model. Our methodology is more general. In fact, differently from the procedure of \cite{BAI2008304} our D$^2$FMs needn't to assume a specific form of nonlinearity either in the encoding or in the decoding map.\footnote{The deep learning literature refers to this as a shift from `feature engineering' to `architecture engineering' \citep[see, for example,][]{PyTorch2019}.}

Finally, we connect with the strand of the deep learning literature that has explored different approaches to introduce dynamics in autoencoders. The early work of \cite{pmlr-v32-gregor14} proposed the Deep AutoRegressive networks (DARN) in which hidden layers are equipped with autoregressive connections, allowing for dynamics in an autoencoder setting. Successively, Temporal Difference Variation Autoencoders (TD-VAEs) were introduced by \cite{gregor2018temporal} to model dynamics in autoencoders via long short-term memory networks (LSTM) connections between belief distributions at two distant time steps. A different line of research in this literature has used Deep Learning for State Space models. For example, \cite{Krishnan2017} adopt MLPs to estimate the mean and covariance matrix of a state space with Gaussian transition dynamics. In \cite{fraccaro2017disentangled} a Kalman Variational Autoencoder (K-VAE) is introduced to estimate (locally) linear Gaussian state space models (LGSSM) by disentangling the observations and the latent dynamics.\footnote{\cite{NIPS2016_6379} uses Structured Variational Autoencoders (SVAEs) to provide conjugate graphical models} Closer to our approach are the Deep State Space Models (DSSM) of \cite{rangapuram2018deep} that directly estimate all the state space parameters using a Recurrent Neural Network (RNN) structure that, given the input features, provides both the latent states and all the time-varying parameters of the state space model. This modelling approach can handle both the presence of noise and missing data. While using deep learning techniques to capture salient features of the data generating process, we tackle those two issues differently. For the former, we apply a denoising approach. For the latter, selection matrices are used to mask missing data when computing the objective to be optimised, while the generative spirit of our model allows to do sampling conditional on the data to fill missing inputs. Importantly, in all the aforementioned approaches, variable specific idiosyncratic components and mixed frequencies are not taken into consideration. We propose a framework able to deal with those two additional data issues. Specifically, we include restriction matrices to aggregate the high frequency latent states to the low frequency as in \cite{mariano2003new} to deal with mixed-frequency, and we design a sequential and iterative (alternated) optimisation scheme between common factors and idiosyncratic components based on a Markov Chain Monte Carlo algorithm.

More generally, deep learning methods have seen early applications in Finance and Economics. In Finance, they have been employed to predict asset prices, stock returns or commodity prices \citep[see][for an extensive literature review]{sezer2019financial}. Closer to our approach, a few recent works have applied neural net to macroeconomic questions. \cite{cook2017macroeconomic} employed a number of neural network architectures, including also autoencoder, to forecast the US unemployment rate. \cite{loermann2019nowcasting} proposed a neural net model to predict the US GDP. \cite{holopainen2017toward} proposed an horse race among different machine learning methods and showed that such models are able to outperform conventional statistical approaches in predicting crisis periods. Finally,  \cite{heaton2016deep}, \cite{gu2018empirical,gu2019autoencoder} employ rich datasets incorporating both stock data and macroeconomic aggregates  to predict stock returns.\\

\section{Autoencoders and Factor Models}\label{sec:background}

Dynamic factor models for econometric times series are multivariate probabilistic models in which a vector of stochastic disturbances are transmitted via linear dynamic equations to the observed variables. They assume that a small number of stochastic unobserved common factors informs the comovements of hundreds of economic variables. In doing so, they combine two core ideas of macroeconomics: the Frisch-Slutsky paradigm that assumes the economic variables to be generated by the stochastic components (the economic shocks) via usually linear dynamic difference equations; and the idea that has guided macro since \cite{RePEc:nbr:nberbk:burn46-1} that a few common disturbances explain most of the dynamics of all the macroeconomics variables, with a residual share due to idiosyncratic components. DFMs are similar in intuition to principal component analysis (PCA) but assume stochastic and dynamic structure that allows their application to econometric time series.

Autoencoders (AE) are neural networks trained to map a set of variables into themselves, by first coding the input into a lower dimensional (or undercomplete) representation) and then decoding it back into itself. The lower dimensional representation forces the autoencoder to capture the most salient features of the data. In constructing a nonlinear reflexive map that links the inputs back to itself via a lower dimensional space, autoencoders can be thought of as a nonlinear generalisation of PCA.\footnote{The connection between PCA and Autoencoders is discussed in \cite{goodfellow2016deep} and in the references therein.} 

In this section, we explore the deep connection between factor models and autoencoders to show that a dynamic formulation of autoencoders can be thought of as a nonlinear generalisations of dynamic factor models, in the same way in which standard autoencoders can be seen as generalisations of principal component analysis. 

\subsection{Latent Factor Models}

Let us first introduce a general formulation of latent factor models with idiosyncratic components. We define $\boldsymbol{y}_t = (y_{t,1},...,y_{t,n})$ as the vector collecting the $n$ variables of interest at time $t$, usually assumed to be the realisation of a vector stochastic process.  A very general latent factor model can be written as
\begin{equation} 
	\boldsymbol{y}_t = F(\boldsymbol{f}_{t}) + \boldsymbol{\varepsilon}_{t} = \widetilde{\boldsymbol{y}}_t + \boldsymbol{\varepsilon}_{t} \label{eq:dfm_classic} \ ,
\end{equation}
where $\boldsymbol{f}_t$ is an $r \times 1$ (for $r=dim(\boldsymbol{f})<<n=dim(\boldsymbol{y})$) vector of latent common stochastic components  -- i.e. the factors --, $\boldsymbol{\varepsilon}_{t}$ are idiosyncratic 
and unobserved stochastic error terms, and $F(\cdot)$ is a generic function mapping the unobserved factors into the observed variable. Usual assumptions are that $\boldsymbol{f}_t$ and $\boldsymbol{\varepsilon}_{t}$ are independent, with zero mean and finite variance (the variance of $\boldsymbol{f}_t$ is often assumed to be a diagonal matrix). For later reference, we indicate as $\widetilde{\boldsymbol{y}}_t$ the component of $\boldsymbol{y}_t$ that relates to the common factors. By assuming also a linear function  $F(\cdot)$, the model reduces to the standard linear factor model
\begin{equation} 
	\boldsymbol{y}_{t} = \boldsymbol{\Lambda} \boldsymbol{f}_{t} + \boldsymbol{\varepsilon}_{t} 
\end{equation}

However, in general, $F(\cdot)$ needn't be linear and we can express the factor component of the model as 
\begin{equation}
	\widetilde{\boldsymbol{y}}_t = F(G(\boldsymbol{y}_t)) = (F \circ G)(\boldsymbol{y}_t)  = (F \circ G)(\widetilde{\boldsymbol{y}}_t + \boldsymbol{\varepsilon}_{t}), \label{eq:gen_encoder}
\end{equation}
where $G(\cdot)$ is the function mapping the observables into the `code' $\boldsymbol{f}_{t}$ (encoding function), and $F(\cdot)$ is the function mapping the factors back into $\boldsymbol{y}_t$ (decoding function). In this form, the connection between factor models and autoencoders is more evident. In fact, the map in Equation \eqref{eq:gen_encoder} can be seen as a very general autoencoder. Linear factor models can be seen as a special case of factor models assuming both a linear encoding and a linear decoding function. It is worth observing that the model in Equations \eqref{eq:dfm_classic} and \eqref{eq:gen_encoder}, without specifying dynamic equations for the stochastic components, can be seen as purely static model.

\subsection{Autoencoders}\label{sec:nn_ae}

Autoencoders belong to the deep neural net family of models and have been introduced for applications involving dimensionality reduction (\citealp{lecun1987phd}, \citealp{bourlard1988auto}, \citealp{hinton1994autoencoders}, \citealp{hinton2006reducing}). Autoencoders solve the parametric problem of finding a mapping (or `learning a representation' in the DNN jargon) of the form $\widetilde{\boldsymbol{y}}_t = F(G(\boldsymbol{y}_t))$ under the constraint of minimising a loss function of choice
\begin{align}
	\mathcal{L}(\boldsymbol{y}_t,\widetilde{\boldsymbol{y}}_t; \boldsymbol{\theta}) = \mathcal{L}(\boldsymbol{y}_t,  F(G(\boldsymbol{y}_t))) \ ,
\end{align}
where $ \mathcal{L}(\cdot)$ is  the loss function and $ \boldsymbol{\theta}$ is the vector collecting all the parameters in $G(\cdot)$ and $F(\cdot)$.

\begin{figure}[t!]
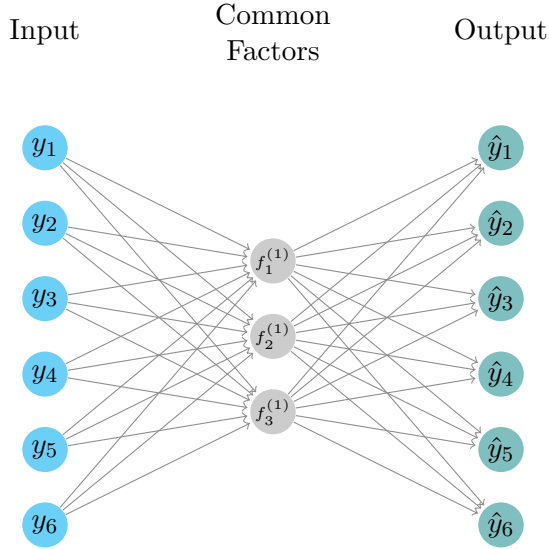

	\centering
	\begin{neuralnetwork}[height=7, layerspacing=30mm]
		\small
		\newcommand{\x}[2]{$y_#2$}
		\newcommand{\y}[2]{$\hat{y}_#2$}
		\newcommand{\hfirst}[2]{\tiny $f^{(1)}_#2$}
		\inputlayer[count=6, bias=false, title=Input, text=\x]
		\hiddenlayer[count=3, bias=false, title=Common Factors, text=\hfirst] \linklayers
		\outputlayer[count=6, title=Output, text=\y] \linklayers					
	\end{neuralnetwork}
	\caption{Principal component analysis (PCA) as an autoencoder.}
	\label{fig:pca_dnn}
\end{figure}

The Principal Components Analysis (PCA) can be seen as the autoencoder minimising the square loss function 
\begin{align}
	\mathcal{L}(\boldsymbol{y}_t,\widetilde{\boldsymbol{y}}_t; \boldsymbol{\theta})  = || \boldsymbol{y}_t- \widetilde{\boldsymbol{y}}_t||^2, 
\end{align}
and assuming both a linear coding and a linear decoding function, i.e.  $\boldsymbol{f}_{t} = G(\boldsymbol{y}_{t}) = W'\boldsymbol{y}_t $ and $\widetilde{\boldsymbol{y}}_t = F(\boldsymbol{f}_{t}) = \boldsymbol{\Lambda} \boldsymbol{f}_{t}$. Figure \ref{fig:pca_dnn} provides a neural net representation of PCA.

\begin{figure}[t!]
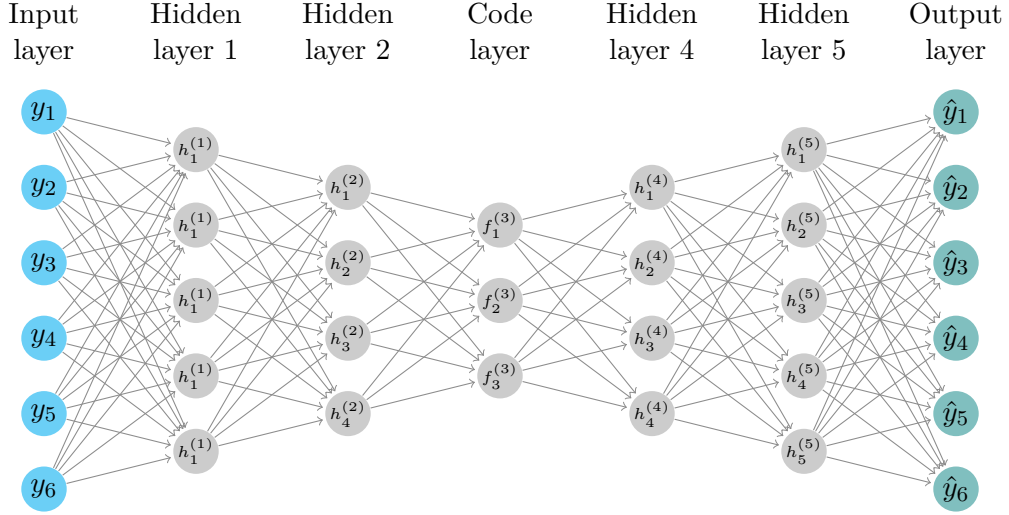
 
	\centering
	\begin{neuralnetwork}[height=6, layerspacing=20mm]
		\small
		\newcommand{\x}[2]{$y_#2$}
		\newcommand{\y}[2]{$\hat{y}_#2$}
		\newcommand{\hfirst}[2]{\tiny $h^{(1)}_#1$}
		\newcommand{\hsecond}[2]{\tiny $h^{(2)}_#2$}
		\newcommand{\hthird}[2]{\tiny $f^{(3)}_#2$}
		\newcommand{\hfourth}[2]{\tiny $h^{(4)}_#2$}
		\newcommand{\hfifth}[2]{\tiny $h^{(5)}_#2$}
		\inputlayer[count=6, bias=false, title=Input\\layer, text=\x]
		\hiddenlayer[count=5, bias=false, title=Hidden\\layer 1, text=\hfirst] \linklayers
		\hiddenlayer[count=4, bias=false, title=Hidden\\layer 2, text=\hsecond] \linklayers
		\hiddenlayer[count=3, bias=false, title=Code layer, text=\hthird] \linklayers
		\hiddenlayer[count=4, bias=false, title=Hidden\\layer 4, text=\hfourth] \linklayers
		\hiddenlayer[count=5, bias=false, title=Hidden\\layer 5, text=\hfifth] \linklayers
		\outputlayer[count=6, title=Output\\layer, text=\y] \linklayers
	\end{neuralnetwork}
	\caption{A symmetric autoencoder with six observables and three neurons in the code layer (biases are not included in the graph). The first two hidden layers operate the encoding, while the last two hidden layers decode into the output.}
	\label{fig:autoenc_classic}
\end{figure}

In principle, $G(\cdot)$ and $F(\cdot)$ can be any nonlinear function and hence finding the correct functional form capturing a data generating process of interest can be a daunting problem. Autoencoders provide a practical implementation of this problem by expressing the composition of two functions as a chain of two multilayer perceptrons (MLPs): the first chain operates the coding, while the second produces the decoded output (see a graphical representation of a symmetric autoencoder in Figure \ref{fig:autoenc_classic}). A multilayer perceptron is a type of feedforward artificial neural network composed of a number of `hidden' layers each formed by a number of `nodes' (or `neurons'). Each neuron in each layer receives some inputs from the neurons in the previous layer and outputs to the next layer an activation output, $h_{m_l}^l$. The activation function (or `link function') of each neuron is a nonlinear function parametrised as
\begin{align}
	h_{m_l}^l = g_{m_l}^l(\boldsymbol{W}_{m_l}^l \boldsymbol{h}^{l-1} + b_{m_l}^l) \ ,
\end{align}
where $l$ is the layer (for $l = 1, \dots, L$), $m_l$ is the node, and $\boldsymbol{\theta}_{m_l}^l \equiv \{\boldsymbol{W}_{m_l}^l, b_{m_l}^l\}$ are the parameters of the activation function to be determined, in the form of, respectively, a set of weights and a constant (also called bias). If $l=1$ then, $\boldsymbol{h}_{l-1} = \boldsymbol{h}_0 = \boldsymbol{y}_t$ that is the input data vector. Common choices for the activation function $g_{m_l}^l(\cdot)$ are the sigmoid, the hyperbolic tangent (tanh), softplus, and the rectified linear unit (ReLu) functions. Hence, in other words, the neuron in each upper layer is the product of an element wise (usually monotone) transformation applied on an affine mapping of the neurons in the lower layer. 

We can define as $\boldsymbol{g_l}(\cdot)$ the vector containing all the activation functions of a layer $\{g_{1}^l(\cdot) \dots g_{M}^l(\cdot)\}'$ for all the nodes $1, \dots M$. Hence the first MLP will be given by the composition of the activation functions of each node in each layer of the encoding feedforward network, i.e.
\begin{align}
	\boldsymbol{f}_{t} = G(\boldsymbol{y}_{t}) = {\boldsymbol{g_L}}(\boldsymbol{g_{L-1}}(\dots({\boldsymbol{g_1}}(\boldsymbol{y}_t)))) \ .
\end{align}
In a similar way it is usually also defined the MLP operating as decoding network, i.e. as a sequence of layer each containing neurons operating as activation functions over a weighted sum of the inputs plus a constant, i.e.
\begin{align}
	\widetilde{\boldsymbol{y}}_t = F(\boldsymbol{y}_{t}) = \boldsymbol{\tilde g_{L'}}(\boldsymbol{\tilde g_{L'-1}}(\dots({\boldsymbol{\tilde g_1}}(\boldsymbol{f}_{t})))) \ ,
\end{align}
where $\tilde g_{L'}$ is the vector of link functions, and $L'$ is the number of hidden layers in the decoding network.

While the functional form adopted for the activation functions may seem arbitrary, yet a network with such a structure can approximate any nonlinear continuous function. In fact, the Universal Approximation Theorem, a key result in the neural net literature, states that a feed-forward network even with a single hidden layer, containing a finite number of neurons, can approximate continuous functions on compact subsets of $\mathbb R^n$, under mild assumptions on the activation function. However, it does not guarantee that the algorithm adopted to estimate the network can learn the correct parameters  (see \citealp{cybenko1989approximation}, \citealp{HORNIK1989359,HORNIK1990551} and \citealp{lu2017expressive}).

The autoencoder is said to be symmetric when the number of hidden layers in the encoding and in the decoding networks are the same (i.e. $L'=L$), otherwise asymmetric. Asymmetric autoencoders usually have  several layers of encoding but only a single layer of decoding (i.e. $L'=1$), hence the decoding network is not a MLP but an SLP (single layer perceptron).\footnote{Asymmetric autoencoders were introduced by \cite{majumdar2017asymmetric} and have been found to be more accurate compared to traditional symmetrically autoencoders for classification accuracy and also to yield slightly better results on compression problems (over the following datasets: MNIST, CIFAR-10, SVHN and CIFAR-100). \cite{majumdar2017asymmetric} argue that the asymmetric structure helps to reduce the number of parameters to estimate and hence the potential extent of overfitting.}

For a given choice of the link functions, the parameter vector of the autoencoder $ \boldsymbol{\theta}$ contains the full set of weights and constants (biases) that define the affine transformations operated by each neuron before the link function is applied. These parameters are determined by minimising the loss function $\mathcal{L}(\boldsymbol{y}_t,\widetilde{\boldsymbol{y}}_t; \boldsymbol{\theta}) = \mathcal{L}(\boldsymbol{y}_t, F(G(\boldsymbol{y}_t)); \boldsymbol{\theta})$, via back-propagation (\citealp{rumelhart1986learning} and \citealp{lecun1987phd}).\footnote{Stochastic gradient descent (SGD) algorithms, proposed by \cite{kiefer1952stochastic}, are commonly adopted in the deep learning literature, and update the gradient of each parameter using only randomly selected subsamples of the training dataset. These subsamples are called `minibatches' and they are equal partition of the original training datasets. The computational cost of SGD algorithms is independent with respect to the sample size. All the optimisation algorithms tend to analyse the training dataset multiple times in order to reach a better estimation of the parameters that relies less on the starting point. A run of the algorithm over the entire dataset is called `epoch'. Common optimisation algorithms are momentum algorithms, as AdaGrad by \cite{duchi2011adaptive}, RMSProp and its variation by \cite{tieleman2012rmsprop}, and ADAM by \cite{Adam}, that rely all on the well know gradient descent algorithm by  \cite{cauchy1847methode}.} 
One way to estimate autoencoders is by corrupting the inputs with noise injection \citep[see][]{vincent2008extracting} during the training process. Those are referred as Denoising Autoencoders. The intuition for this procedure is that, as discussed in \cite{vincent2008extracting} and \cite{bengio2013generalized}, it forces the model to learn the data distribution and not only the distribution specific of the sample used, thanks to data augmentation. Also, noise injection can be seen as a procedure to improve the robustness of neural networks. 

\subsection{Dynamics in Factor Models}\label{sec:linear_dfm}

So far we have discussed the general structure of factor models by abstracting from the dynamics and focusing on the `static' map into lower dimensional factors. Dynamics is usually introduced in DFMs by assuming that both $\boldsymbol{f}_t$ and $\boldsymbol{\varepsilon}_{t}$ are generated by linear stochastic vector difference equations. For example, \cite{banbura2010nowcasting} and \cite{banbura2014maximum} consider a system specified as
\begin{align}
	\boldsymbol{y}_{t} &= \boldsymbol{\Lambda} \boldsymbol{f}_{t} + \boldsymbol{\varepsilon}_{t} \ 
	\label{eq:eq_meas_a}
	, \\ 
	\boldsymbol{f}_t &= \boldsymbol{B}_1\boldsymbol{f}_{t-1} + \dots + \boldsymbol{B}_p\boldsymbol{f}_{t-p} + \boldsymbol{u}_t, \quad \quad  \boldsymbol{u}_t \iidN(0,\boldsymbol{U}), \label{eq:eq_fact} \\
	\boldsymbol{\varepsilon}_t &= \boldsymbol{\Phi}_1 \boldsymbol{\varepsilon}_{t-1} + \dots + \boldsymbol{\Phi}_d \boldsymbol{\varepsilon}_{t-d} + \boldsymbol{\epsilon}_t , \quad  \quad \boldsymbol{\epsilon}_t \iidN(0, \boldsymbol{Q}) , \label{eq:eq_idio_a}
\end{align}
where $\boldsymbol{B}_1, \dots, \boldsymbol{B}_p$ are the $r \times r$ matrices of autoregressive coefficients for the factor and $\boldsymbol{\Phi}_1, \dots, \boldsymbol{\Phi}_d$ are the $n \times n$ diagonal matrices of autoregressive coefficients for the idiosyncratic component (i.e. $\boldsymbol{\Phi}_1 = diag(\phi_1, \dots, \phi_n)$). Specifically, \cite{banbura2010nowcasting} and \cite{banbura2014maximum} assume a VAR process of order two ($p= 2$) for factors, and of order one ($d= 1$) for the idiosyncratic component.\footnote{The zero cross-correlation at all leads and lags of the idiosyncratic components has been shown by \cite{doz2012quasi} and \cite{barigozzi2019quasi} to be asymptotically valid even when it is mildly violated in small sample.}

Such a structure can be seen, in our framework, as obtained by assuming that:
\begin{itemize}
	\item[A.1] {\bf Encoding function} $G_{\theta_G}(\cdot): \boldsymbol{y} \rightarrow \boldsymbol{f}$ is a linear operator;
	\item[A.2] {\bf Decoding function} $F_{\theta_F}(\cdot): \boldsymbol{f} \rightarrow \widetilde{\boldsymbol{y}}$ is a linear operator;
	\item[A.3] {\bf Factor dynamics} $\boldsymbol{f}_{t}$ follows a linear stochastic vector difference equation;
	\item[A.4] {\bf Idiosyncratic component dynamics} $\boldsymbol{\varepsilon}_{t}$ follows a linear stochastic vector difference equation with diagonal matrices of autoregressive coefficients;
	\item[A.5] {\bf Distributions} Error terms from the transition (and emission) equations are assumed to be $i.i.d.$ Gaussian.\footnote{Once in state-space, a standard DFM as described in equations from \eqref{eq:eq_meas_a} to \eqref{eq:eq_idio_a} features a noise process in the measurement equation \eqref{eq:eq_meas_a}, on top of the error terms $\boldsymbol{u}_t$ and $\boldsymbol{\epsilon}_t$ from the transition equations. This measurement error term, $\boldsymbol{\eta}_t$, is usually assumed to be $i.i.d.$ multivariate Gaussian with identity matrix scaled by a small constant, that is $\boldsymbol{\eta}_t \iidN(0, \eta \boldsymbol{I})$, with $\eta$ a small number larger than zero.}
\end{itemize}

Autoencoders provide a practical solution to estimate  factor models with a more general structure, potentially relaxing one or all of these assumptions to obtain both nonlinear maps from reduced dimension factors to variables and vice-versa, but also to introduce nonlinear dynamic equations. This approach to the generalisation of dynamic factor models, is what we call Deep Dynamic Factor Models (D$^2$FMs). In the next section, we show how to construct and estimate an autoencoder that relaxes assumptions A.1 and A.2, while maintaining the others.\footnote{See \cite{doz2012quasi} for the robusteness of their DFM estimation procedure with respect to assumptions A.4 and A.5.}

\subsection{Estimation and Conditional Likelihood}\label{sec:ML}

In principle the parameters of a parametric factor model of the form $\boldsymbol{y}_t = F(\boldsymbol{f}_{t}) + \boldsymbol{\varepsilon}_{t} $ would be estimated via maximum likelihood,
\begin{equation}
	\boldsymbol{\widehat \theta}  = \argmax_{\boldsymbol{\theta}} p_{model}(\boldsymbol{Y}; \boldsymbol{\theta}) \ ,
\end{equation}
where by $\boldsymbol{Y}$ is the full sample of observation, and $p_{model}(\cdot; \cdot) $ is the conditional probability density function of the model.

However, a direct maximum likelihood is rarely feasible, even for linear models, and iterative methods to find maximum likelihood or maximum a posteriori (MAP) estimates of the parameters of the model are preferred. In fact, maximum likelihood estimators of the parameters $\boldsymbol{\theta} = (\boldsymbol{\Lambda}, \boldsymbol{B}(L), \boldsymbol{U}, \boldsymbol{\Phi}(L), \boldsymbol{Q})$ are in general not available in closed form and a direct numerical maximisation can be too demanding when $n$ is large. 
Indeed, a proposed solution in the linear factor model literature is to adopt the Expectation Maximisation (EM) algorithm, a maximum a posteriori method, and to initialise the common factors $\boldsymbol{f}_t$ with PCA on the observables.\footnote{Estimation of linear factor models was originally carried out via simple principal component analysis (PCA).} The updates of the latent components are performed using the Kalman filter and smoother. 

A similar approach can be adopted from a `deep' point of view on factor models by employing the methodologies developed in the deep learning literature, without losing the dynamic model interpretation. As we discuss in the next section, the model parameters of a D$^2$FM can be estimated via Monte Carlo gradient methods, instead of using the EM algorithm. This has computational advantages -- the methods are fast and reliable -- even when the dataset is big. At the same time, estimation results can be thought of as approximating a  maximum a posteriori method.

It is well know, in the linear case, that if the innovation $\boldsymbol{\varepsilon}_{t}$ are assumed to be independent (or uncorrelated) of $\boldsymbol{f}_{t}$ and normally distributed, than the maximisation of the likelihood with respect to the parameters of the model yields the same estimate for the parameters as does minimising the mean squared error. 

Importantly, this equivalence between maximum likelihood estimation and minimisation of mean squared error holds regardless of the function used to predict the mean of the Gaussian distributed variable $\boldsymbol{y}_t$ \citep[see][]{goodfellow2016deep}. This result allows for an interpretation of estimation results from autoencoders with mean squared error from a Bayesian perspective, using standard likelihood methods, or from a frequentist one as the (approximated) mean estimator of a Gaussian distributed process.

Furthermore, the equivalence between maximum likelihood estimation and minimisation of mean squared error together with the Universal Approximation Theorem allow to reinterpret the autoencoders and the procedure adopted in estimating them as an efficient computational method to approximate the maximum likelihood estimates of nonlinear factor models. These are dynamic models that are defined by a conditionally Gaussian distribution centred around a mean provided by a nonlinear but continuous function of the inputs. In the next section, we provide an algorithm that implements Deep Dynamic Factor Models.

\section{D$^2$FM  Estimation}\label{sec:estimation_sec}

In its general form, the D$^2$FM can be written as
\begin{align}
	\boldsymbol{f}_{t} &= G(\boldsymbol{y}_{t}) \ , \label{eq:encoding}\\
	\boldsymbol{y}_{t} &= F(\boldsymbol{f}_{t}) + \boldsymbol{\varepsilon}_{t} \ ,  \label{eq:decoding}\\ 
	\boldsymbol{f}_t &= \boldsymbol{B}_1\boldsymbol{f}_{t-1} + \dots + \boldsymbol{B}_p\boldsymbol{f}_{t-p} + \boldsymbol{u}_t, \quad \quad  \boldsymbol{u}_t \iidN(0,\boldsymbol{U}), \label{eq:factor-dyn}\\
	\boldsymbol{\varepsilon}_t &= \boldsymbol{\Phi}_1 \boldsymbol{\varepsilon}_{t-1} + \dots + \boldsymbol{\Phi}_d \boldsymbol{\varepsilon}_{t-d} + \boldsymbol{\epsilon}_t , \quad  \quad \boldsymbol{\epsilon}_t \iidN(0, \boldsymbol{Q}), \label{eq:idio-dyn}
\end{align}
where the assumptions on the linearity of the dynamic equations are maintained, Equations \eqref{eq:factor-dyn} and \eqref{eq:idio-dyn}, while the model allows for a nonlinear map between variables and factors.\footnote{Alternatively, the dynamic of the common latent states can be estimated directly in the algorithm \ref{algo:our_algo}. This can be achieved by including an autoregressive layer before the decoding layer ($F_{\boldsymbol{\theta}_2}$ of the algorithm). This additional layer would coincide with the state equation of the common part. This layer could be linear for linear dynamics, but also nonlinear and composed of multiple layers such as multi-layer perceptrons or recurrent layers such as LSTM.} The estimation of linear factor dynamics separately creates what \cite{stock2011dynamic} call a `state space with static (common) factors' as opposed to a `state space with dynamic (common) factors'.

The D$^2$FM can be implemented using a symmetric autoencoder architecture with a MLP capturing the encoding function, Equation \eqref{eq:encoding}, and another MLP providing the nonlinear decoding, Equation \eqref{eq:decoding}. The assumption of i.i.d. and Gaussian innovations allows for an interpretation of the estimated network as MAP of the likelihood of the model \citep[see][]{goodfellow2016deep}. 

Importantly, such a model specification encompasses several simplified models, most notably the standard linear DFMs, and hence the estimation algorithm can be specialised to the scope. 

\subsection{Network Design}\label{sec:networkdesign}

\begin{figure}[t!]
	\centering
	\includegraphics[width=1\textwidth]{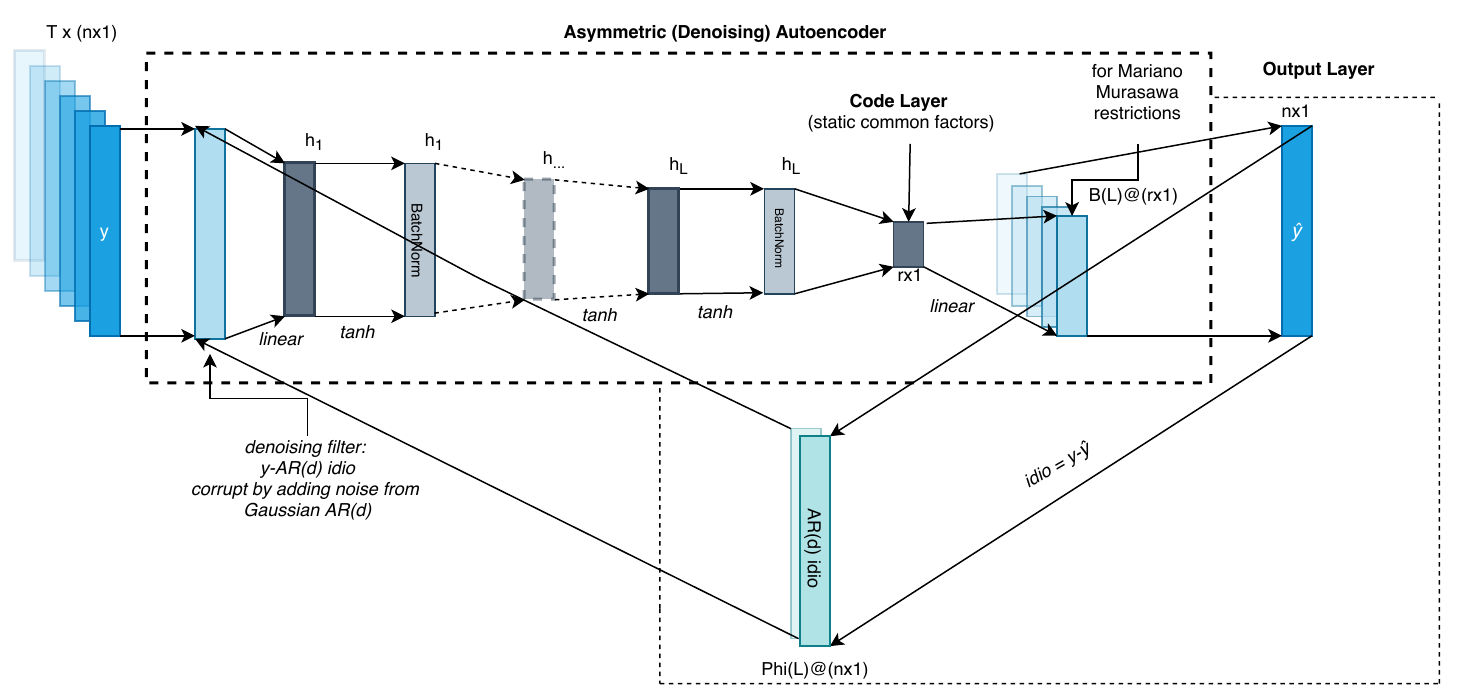}
	\caption{A graph representation of the training process for a the D$^2$FM with an asymmetric structure: nonlinear multilayer encoder and linear single layer decoder.}
	\label{fig:model_graph} 
\end{figure}

The core of the model is provided by an autoencoder with a nonlinear multilayer encoder and either a symmetric structure in the decoding, for nonlinear decoding, or an asymmetric structure with a linear single layer decoder. Linear stochastic autoregressive equations are adopted to model the dynamics of factors and idiosyncratic components. Figure \ref{fig:model_graph} shows a diagrammatic representation of the model with a single layer decoder.

The number of hidden layers in the encoding network as well as the number of neurons need not be pre-specified but can be selected via cross-validation. This enables the model to capture different degrees of complexity in the data, which is not know a-priori. With respect to the activation functions, we equip each neuron in the coding layers with a link function in the form of the hyperbolic tangent (\textit{tanh}) for the real-time macroeconomic dataset, and of the rectified linear unit (ReLU) for the Monte Carlo exercises.\footnote{In general, some tuning might be needed in order to find the correct specification for the dataset at use.} In the encoding multilayer perceptron we also include two batch normalisation layers to induce some regularisation, and control over potential covariate shift \citep[see][]{pmlr-v37-ioffe15}.\footnote{Covariate shift is a phenomenon that occurs in deep learning when the distribution of the input data changes between the training and testing phases. This can lead to a decrease in the accuracy of the model because the model has not been exposed to the new distribution during training.} 

In the decoding network, an additional linear layer can be included to introduce constraints needed to account for the mixed frequencies of macroeconomic data. This additional layer does not have any additional parameter, and it only includes aggregation weights to map the high frequency latent states to the low frequency observables.

\subsection{Estimation and Online Learning of the D$^2$FM}\label{sec:estimation}

In our estimation of the D$^2$FM, we propose a two-step procedure to differentiate between on-line (out-of-sample) and off-line (in-sample) learning.\footnote{In the deep learning literature, online learning means that the model estimates parameters using the flow of data as it comes in (or using a simulated flow). Offline means that the model employs a static dataset. This is similar to the standard econometric distinction between out-of-sample and in-sample.} 
\begin{itemize}
	\item[-] {\bf Step 1} estimate off-line all the parameters of the model;
	\item[-] {\bf Step 2} cast the decoding part in a state-space framework to allow for on-line updates of the latent states given the observables.
\end{itemize}

Algorithm \ref{algo:our_algo} implements the off-line estimation step ({\bf Step 1}) of our D$^2$FM, assuming an AR(d) for $p_{idio}(\cdot)$, but possibly a general encoding $G_{\theta_G}(\cdot)$ and decoding $F_{\theta_F}(\cdot)$ function. The proposed algorithm for estimating D$^2$FM builds on and extends what has been proposed by \cite{bengio2013generalized} to estimate Generalised Denoising Autoencoders.

\begin{algorithm}
	\caption{MCMC for D$^2$FM with stationary AR(d) idiosyncratic components -- \small \textit{requires a training set, an encoding structure $G_{\boldsymbol{\theta}_G}(\cdot)$ and a decoding one $F_{\boldsymbol{\theta}_F}(\cdot)$}} 
	\label{algo:our_algo}
	\begin{algorithmic}[1]
		\Statex \textbf{init:} $\boldsymbol{\theta}_G, \boldsymbol{\theta}_F,\boldsymbol{\Phi}, \Sigma_{\varepsilon}, \boldsymbol{\varepsilon}_{t}$
		\Statex \textbf{repeat}
		\smallskip
		\State $\quad \widetilde{\boldsymbol{y}}_{t} | (\boldsymbol{y}_{t}, \hat{\boldsymbol{\varepsilon}}_t) =  \boldsymbol{y}_{t} - \boldsymbol{\Phi}(L) \boldsymbol{\varepsilon}_{t}$
		\smallskip
		\State $\quad$ \textbf{Loop} \text{epochs, batches} \textbf{Do}
		\smallskip
		\State $\quad \quad$draw $\boldsymbol{\varepsilon}^{(mc)}_{t} \iidN(0, \Sigma_{\varepsilon})$
		\smallskip
		\State $\quad \quad \boldsymbol{y}^{(mc)}_{t} = \widetilde{\boldsymbol{y}}_{t} | (\boldsymbol{y}_{t}, \hat{\boldsymbol{\varepsilon}}_t)  + \boldsymbol{\varepsilon}^{(mc)}_{t}$
		\smallskip
		\State $\quad \quad \theta_G, \theta_F$ update by a gradient based step on $\mathcal{\hat{L}}(\boldsymbol{y}_t,F_{\boldsymbol{\theta}_F}(G_{\boldsymbol{\theta}_G}(\boldsymbol{y}^{(mc)}_{t})))$
		\smallskip
		\State $\quad$ \textbf{End Loop}
		\smallskip
		\State $\quad \boldsymbol{f}_t | \boldsymbol{y}^{(mc)}_{t}  = \mathbb{E}_{\boldsymbol{y}^{(mc)}_{t} \sim \boldsymbol{y}_{t}, \hat{\boldsymbol{\varepsilon}}_t} G_{\boldsymbol{\theta}_G}(\boldsymbol{y}^{(mc)}_{t})$ 
		\smallskip
		\State $\quad \boldsymbol{\varepsilon}_t | \boldsymbol{y}_t, \boldsymbol{f}_t = \boldsymbol{y}_t - F_{\boldsymbol{\theta}_F}(\boldsymbol{f}_t | \boldsymbol{y}^{(mc)}_{t})$
		\smallskip
		\State $\quad \boldsymbol{\Phi} \gets stationary \; AR(d) \;$ \text{on} $\;\boldsymbol{\varepsilon}_t$
		\smallskip
		\State $\quad \Sigma_{\varepsilon} \gets from \; \boldsymbol{\varepsilon}_t$
		\smallskip
		\Statex \textbf{until} convergence on $\mathcal{\hat{L}}(\boldsymbol{y}_t, F_{\boldsymbol{\theta}_F}(\boldsymbol{f}_t | \boldsymbol{y}^{(mc)}_{t}))$ in $L_1$ norm
		\smallskip
		\Statex \textbf{return} $\Sigma_{\varepsilon}, \boldsymbol{\Phi}, \boldsymbol{f}_t, F_{\boldsymbol{\theta}_F}$
	\end{algorithmic}
	\label{algo:algo1}
\end{algorithm}

Let us present the estimation algorithm. Parameters are first initialised. Line 1 performs a filtering of the input data ${\boldsymbol{y}}_{t}$ by using the conditional mean of the AR(d) of the idiosyncratic components. From lines 2 to 6, the Monte Carlo step and the gradient updates over each epoch and batch are carried out, employing the filtered data $\widetilde{\boldsymbol{y}}_{t}$ and injecting Gaussian noise from $\boldsymbol{\varepsilon}_{t}$ in a denoising fashion to obtain the noisy observations $\boldsymbol{y}^{(mc)}_{t}$. In line 7, the latent states $\boldsymbol{f}_t$ are extracted from the encoding network via Monte Carlo integration, while from line 8 to 10 the algorithm updates the parameters of the idiosyncratic process $\boldsymbol{\varepsilon}_t$, conditional on the factors and the observables. The adoption of an $L_2$ (MSE) loss function $\mathcal{\hat{L}}(\boldsymbol{y}_t,F_{\boldsymbol{\theta}_F}(\boldsymbol{f}_t | \boldsymbol{y}^{(mc)}_{t}))$ allows for interpretability of the results, as discussed. We specify an estimated loss, as missing data prevents us from deriving the exact loss.\footnote{We give details about the treatment of missing data in Section \ref{sec:miss_data}.} Finally, convergence is checked as the $L_1$ norm of the distance between the loss function at two iterations. It is worth noting that the loss, $\mathcal{\hat{L}}(\cdot)$ includes only the common components, since under our assumptions at convergence we have the following decomposition of the log-likehood: 
\begin{equation}
	\begin{split}
		& \log p_{model}(\boldsymbol{y}_t | \boldsymbol{f}_t = G_{\theta_G}(\boldsymbol{y}^{(mc)}_{t}), \boldsymbol{\varepsilon}_t = \hat{\boldsymbol{\varepsilon}}_t) = \\ &
		\log p_{decoder}(\boldsymbol{y}_t | \boldsymbol{f}_t = G_{\theta_G}(\boldsymbol{y}^{(mc)}_{t})) + \log
		p_{idio}(\boldsymbol{y}_t | \boldsymbol{\varepsilon}_t = \hat{\boldsymbol{\varepsilon}}_t) \ ,
	\end{split}
\end{equation}
where $\hat{\boldsymbol{\varepsilon}}_t$ is the estimated idiosyncratic autoregressive component. In running over epochs and batches (lines 2 to 6), the algorithm injects uncorrelated noise into the data (it is a Denoising Autoencoder). Hence it searches for a maximum a posteriori of the parameters for the modified model with log-likelihood
\begin{equation}\label{eq:NCAE_loglikehood_all}
	\mathbb{E}_{\boldsymbol{y}_{t}\sim\ p_{data}(\boldsymbol{y}_{t})}\mathbb{E}_{\boldsymbol{y}^{(mc)}_{t} \sim p_{noisy}(\boldsymbol{y}^{(mc)}_{t} | \boldsymbol{y}_t, \hat{\boldsymbol{\varepsilon}}_t)} \log \: p_{model}(\boldsymbol{y}_t | \boldsymbol{f}_t = G_{\theta_G}(\boldsymbol{y}^{(mc)}_{t}), \boldsymbol{\varepsilon}_t = \hat{\boldsymbol{\varepsilon}}_t) \ ,
\end{equation}
where $p_{noisy}(\boldsymbol{y}^{(mc)}_{t} | \boldsymbol{y}_t, \hat{\boldsymbol{\varepsilon}}_t)$ is the corruption distribution, using a Gaussian autoregressive process. The idea behind this procedure is to filter out the foreseeable idiosyncratic part from the input variables, so that only the common component(s) remain(s). Injecting noise from the unconditional idiosyncratic distribution will generate new samples which are not unreasonably far from the old ones. In doing so, we define an appealing and convenient linkage between the corruption process of the denoising approach and the idiosyncratic component distribution ($p_{idio}$).

In {\bf Step 2}, the output of the algorithm is cast in the state-space of equations \eqref{eq:decoding}-\eqref{eq:idio-dyn}. Dynamics of the common factors are estimated via OLS or Maximum Likelihood.\footnote{The dynamics of the common latent states can also be estimated directly in Algorithm \ref{algo:our_algo}. This can be achieved by including an additional autoregressive layer before the decoding network ($F_{\boldsymbol{\theta}_F}$ of the algorithm).} State updates can then be carried out via either nonlinear filtering procedures for a nonlinear decoder, or via Kalman filtering in the presence of a linear decoder. This allows for online (i.e., out-of-sample) learning with the flow of data.

\subsection{Cross-Validating Hyperparameters}\label{sec:cv_val}

The D$^2$FM described in this section is subject to the selection of a number of critical parameters determining its structure, beyond the coefficients $\boldsymbol{\theta}$. These parameters are commonly known as hyperparameters, being them set before the training starts and usually selected over a grid with respect to some validation loss, which is estimated via a process called cross-validation.

The D$^2$FM has hyperparameters typical of both deep learning and time-series models. In particular, the deep learning hyperparameters can be divided into two categories. The first relates to the neural network structure and includes: the type of layers, the number of hidden layers, the number of neurons per each hidden layer, penalisation coefficients, dropout layers and relative dropout rates (if included), batch normalisation layers and the link function used. The second category relates to the optimisation algorithm used and comprehends: size of the mini-batches, number of epochs, the learning rate and the momentum coefficients of the gradient optimisation method, if present.  Standard time-series factor models have few additional hyperparameters which include: the number of latent common states, the number of lags of the input variables, the number of lags of the latent common states and of the idiosyncratic states. These hyperparameters, in the time series literature, are either fixed a-priori or estimated using information criteria instead of grid-search algorithms.\footnote{The Akaike information criteria (AIC) and the Bayesian information criteria (BIC) can be used to determine the number of lags; while the number of latent factors can be, in principle, estimated using the method proposed by \cite{alessi2010improved} which improves \cite{bai2002determining}'s methodology.}

It is important to observe that in time-series we cannot apply the common `K-fold' cross-validation method because usually its technical conditions are not met \citep[see][for details]{bergmeir2018note}. Therefore, we use a standard out-of-sample approach which consists in splitting the set of observations available up to a certain point in time, $T$, between a training set $[0, T-k*h-1]$, and validation set $[T-k*h, T-(k-1)*h]$, where $h$ determines the length of the set, while $k = K,\dots,1$ with $K \ll \frac{T-1}{h}$. By averaging over the losses computed on the $K$ `validation sets', we get an estimate of the `validation loss'. This means that we need to estimate a given model with fixed hyperparameter $K$ times, and this for each possible hyperparameters combination. Therefore, with deterministic search method the computational cost is exponential in the dimensionality of the hyperparameters.\footnote{Alternative methods based on stochastic search are available (see for example \citealp{bergstra_tpe}), but then the results could be not robust when they are computed on a small number of iterations.}

\section{Monte Carlo Experiment}\label{sec:mc_sim}

In this section, we compare the performances of DFMs and of D$^2$FM in a controlled environment, by simulating artificial time series data with Monte Carlo experiments. In doing so, we first consider the linear data generating processes studied in \cite{doz2012quasi} and \cite{banbura2014maximum}, and then, adopt its nonlinear generalisation proposed by \cite{gu2019autoencoder}. In particular, we consider data generated the following process:
\begin{align}
	\boldsymbol{y}_{t} &= F(\boldsymbol{f}_{t}) + \boldsymbol{\varepsilon}_{t} + \boldsymbol{v}_t \ ,  \\ 
	\boldsymbol{f}_t &= \boldsymbol{B}_1\boldsymbol{f}_{t-1} + \boldsymbol{u}_t, \quad \quad  \boldsymbol{u}_t \iidN(0,\boldsymbol{I}_{r}), \\
	\boldsymbol{\varepsilon}_t &= \boldsymbol{\Phi}_1 \boldsymbol{\varepsilon}_{t-1} + \boldsymbol{\epsilon}_t , \quad  \quad \boldsymbol{\epsilon}_t \iidN(0, \boldsymbol{Q}), 
\end{align}
for $t=1,\dots,T$. As in \cite{gu2019autoencoder}, we simulate both a linear, and nonlinear factor model, and allow $F(\cdot)$ to take two forms:
\begin{equation}
	F(\boldsymbol{f}_{t}) = 
	\begin{cases}
		\boldsymbol{\Lambda} \boldsymbol{f}_t & \text{if linear} \\ 
		\boldsymbol{\Lambda} [\boldsymbol{f}_t, poly(\boldsymbol{f}_t, 2), sgn(\boldsymbol{f}_t)]' & \text{if nonlinear}
	\end{cases},
\end{equation}
where $sgn(\cdot)$ is the sign function ($1$ if positive, $-1$ if negative) and $poly(\cdot,2)$ is a generator of polynomial functions of order $2$. The parameters of the model are set as follows:
\begin{eqnarray}
	&\boldsymbol{\Lambda}_{ij} \iidN(0,1), \qquad \ i=1,\dots,n, \qquad \, & j=1,\dots,\tilde{r}, \\
	&\tilde{r} =
	\begin{cases}
		r & \text{linear}\\
		\frac{4r + r(r+1)}{2} & \text{nonlinear}
	\end{cases}, \\ 
	&\boldsymbol{B}_{ij,1} = 
	\begin{cases} 
		\rho & \text{if } i=j \\
		0 & \text{otherwise}
	\end{cases},
	& \qquad 
	\boldsymbol{\Phi}_{ij,1} = 
	\begin{cases}
		\alpha & \text{if } i=j \\
		0 & \text{otherwise}
	\end{cases},
	\\
	&\boldsymbol{Q}_{ij} = \tau^{|i-j|}(1-\alpha^2)\sqrt{\gamma_i \gamma_j}, & \quad \gamma_i=\frac{\beta_i}{1-\beta_i}\frac{1}{1-\rho^2} \sum_{j=1}^{\tilde{r}} \boldsymbol{\Lambda}^2_{ij},\\ 
	&\beta_i \sim U([u, 1-u]).
\end{eqnarray}
We consider a range of possible specifications of the parameters by setting the number of factors $r=\{1, 3\}$, the number of variables $n=\{10, 100\}$, autoregressive coefficient of the factors $\rho=\{0.5, 0.9\}$, autoregressive coefficient of the idiosyncratic component $\alpha=\{0, 0.5\}$, the number of observations $T=\{50, 200\}$, and the fraction of missings is in $\{0, 0.3\}$. 

For each setting, we run 100 Monte Carlo simulations and estimate a DFM, and a four layers D$^2$FM, with a ReLU activation function augmented with three \textit{BatchNorm} layers. The number of factors is set for both models to the true number of factors (i.e. $r$ when the DGP is linear in the factors and $\tilde{r}$ when it is nonlinear). Starting from the factor layer, hidden neurons increase by a factor of two in each layer up to the input layer.

As in \cite{stockwatson2002a}, \cite{doz2012quasi} and \cite{banbura2014maximum}, we compare the models based on the trace $R^2$ of the regression of the estimated factors on the true ones
\begin{equation}
	\frac{Trace(F'\hat{F}(\hat{F}'\hat{F})^{-1}\hat{F}'F)}{Trace(F'F)} \ ,
\end{equation}
where $\hat{F}=\mathbb{E}_{\boldsymbol{\hat{\theta}}}[F|H_{T}]$ and $H_{T}$ is the history of the data.

In the linear case, differences in performances between the D$^2$FM and the DFM (Table \ref{tab:mc_sim_l}) are very small, and indicate a marginal gain for one model or the other, depending on the specific case. This points to the fact that the two framework are generally equivalent when the DGP is linear. In other words, the D$^2$FM can be seen as a computational efficient way to estimate linear factor models, conditional on a linear DGP.

In the nonlinear simulations (Table \ref{tab:mc_sim_nl}), the D$^2$FM is instead clearly superior. Strikingly, all the differences are statistically significant and the D$^2$FM can explain between $15\%$ and $34\%$ more of the total variance of the simulated common factors. These results validate the ability of the D$^2$FM to better handle several forms of nonlinearities in the DGP, as compared to a standard DFM. 

\begin{table}[htbp]
	\centering
	\begin{tabular}{|rrrr|rrl|rrl|}
		\toprule
		\multicolumn{4}{|c|}{\textbf{Factors}} & \multicolumn{6}{c|}{1} \\
		\midrule
		\multicolumn{4}{|c|}{\textbf{Sample}}  & \multicolumn{3}{c|}{50} & \multicolumn{3}{c|}{200} \\
		\midrule
		\multicolumn{1}{|c}{$\alpha$} & \multicolumn{1}{c}{$\rho$} & \multicolumn{1}{c}{N vars} & \multicolumn{1}{c|}{Missings} & \multicolumn{1}{c}{D$^2$FM} & \multicolumn{1}{c}{DFM} & \multicolumn{1}{c|}{Diff.} & \multicolumn{1}{c}{D$^2$FM} & \multicolumn{1}{c}{DFM} & \multicolumn{1}{c|}{Diff.} \\
		0     & 0.5   & 10    & 0     & 0.91  & 0.89  & \textbf{0.025***} & 0.94  & 0.94  & -0.008*** \\
		0     & 0.5   & 10    & 0.3   & 0.88  & 0.83  & \textbf{0.045***} & 0.91  & 0.91  & -0.006*** \\
		0     & 0.5   & 100   & 0     & 0.96  & 0.95  & \textbf{0.011***} & 0.99  & 0.99  & -0.001*** \\
		0     & 0.5   & 100   & 0.3   & 0.95  & 0.93  & \textbf{0.02***} & 0.99  & 0.99  & \multicolumn{1}{r|}{\textbf{0.001}} \\
		0     & 0.9   & 10    & 0     & 0.74  & 0.75  & \multicolumn{1}{r|}{-0.011} & 0.94  & 0.95  & -0.01*** \\
		0     & 0.9   & 10    & 0.3   & 0.71  & 0.70  & \textbf{0.001*} & 0.93  & 0.94  & -0.013*** \\
		0     & 0.9   & 100   & 0     & 0.77  & 0.74  & \multicolumn{1}{r|}{\textbf{0.024}} & 0.96  & 0.96  & \textbf{0.001***} \\
		0     & 0.9   & 100   & 0.3   & 0.76  & 0.75  & \textbf{0.017***} & 0.96  & 0.96  & \multicolumn{1}{r|}{\textbf{0.002}} \\
		0.5   & 0.5   & 10    & 0     & 0.90  & 0.85  & \textbf{0.043***} & 0.92  & 0.92  & \multicolumn{1}{r|}{\textbf{0.001}} \\
		0.5   & 0.5   & 10    & 0.3   & 0.85  & 0.77  & \textbf{0.086***} & 0.88  & 0.89  & \multicolumn{1}{r|}{-0.008} \\
		0.5   & 0.5   & 100   & 0     & 0.96  & 0.94  & \textbf{0.013***} & 0.99  & 0.99  & -0.001*** \\
		0.5   & 0.5   & 100   & 0.3   & 0.95  & 0.94  & \textbf{0.015***} & 0.98  & 0.99  & -0.001*** \\
		0.5   & 0.9   & 10    & 0     & 0.72  & 0.72  & \textbf{0.004***} & 0.93  & 0.93  & \multicolumn{1}{r|}{0} \\
		0.5   & 0.9   & 10    & 0.3   & 0.71  & 0.70  & \textbf{0.007***} & 0.92  & 0.93  & -0.004*** \\
		0.5   & 0.9   & 100   & 0     & 0.77  & 0.73  & \textbf{0.035**} & 0.96  & 0.96  & \textbf{0.001***} \\
		0.5   & 0.9   & 100   & 0.3   & 0.76  & 0.75  & \textbf{0.018***} & 0.96  & 0.96  & \textbf{0.002***} \\
		\midrule
		\multicolumn{4}{|c|}{\textbf{Factors}} & \multicolumn{6}{c|}{3} \\
		\midrule
		\multicolumn{4}{|c|}{\textbf{Sample}}  & \multicolumn{3}{c|}{50} & \multicolumn{3}{c|}{200} \\
		\midrule
		\multicolumn{1}{|c}{$\alpha$} & \multicolumn{1}{c}{$\rho$} & \multicolumn{1}{c}{N vars} & \multicolumn{1}{c|}{Missings} & \multicolumn{1}{c}{D$^2$FM} & \multicolumn{1}{c}{DFM} & \multicolumn{1}{c|}{Diff.} & \multicolumn{1}{c}{D$^2$FM} & \multicolumn{1}{c}{DFM} & \multicolumn{1}{c|}{Diff.} \\
		0     & 0.5   & 10    & 0     & 0.71  & 0.71  & -0.001** & 0.76  & 0.80  & -0.039*** \\
		0     & 0.5   & 10    & 0.3   & 0.60  & 0.58  & \multicolumn{1}{r|}{\textbf{0.021}} & 0.66  & 0.71  & -0.051*** \\
		0     & 0.5   & 100   & 0     & 0.94  & 0.91  & \textbf{0.021***} & 0.97  & 0.97  & -0.002*** \\
		0     & 0.5   & 100   & 0.3   & 0.92  & 0.91  & \textbf{0.014***} & 0.96  & 0.96  & \multicolumn{1}{r|}{\textbf{0.002}} \\
		0     & 0.9   & 10    & 0     & 0.63  & 0.66  & -0.029*** & 0.82  & 0.88  & -0.06*** \\
		0     & 0.9   & 10    & 0.3   & 0.58  & 0.64  & -0.066*** & 0.75  & 0.85  & -0.101*** \\
		0     & 0.9   & 100   & 0     & 0.74  & 0.74  & \textbf{0.003***} & 0.92  & 0.92  & -0.001*** \\
		0     & 0.9   & 100   & 0.3   & 0.73  & 0.73  & \multicolumn{1}{r|}{-0.002} & 0.92  & 0.92  & -0.003*** \\
		0.5   & 0.5   & 10    & 0     & 0.67  & 0.63  & \textbf{0.044***} & 0.70  & 0.69  & \multicolumn{1}{r|}{\textbf{0.01}} \\
		0.5   & 0.5   & 10    & 0.3   & 0.56  & 0.52  & \textbf{0.035***} & 0.60  & 0.61  & -0.013*** \\
		0.5   & 0.5   & 100   & 0     & 0.93  & 0.91  & \textbf{0.021***} & 0.97  & 0.97  & \multicolumn{1}{r|}{0} \\
		0.5   & 0.5   & 100   & 0.3   & 0.92  & 0.88  & \textbf{0.033***} & 0.96  & 0.95  & \multicolumn{1}{r|}{\textbf{0.002}} \\
		0.5   & 0.9   & 10    & 0     & 0.60  & 0.63  & -0.031*** & 0.77  & 0.85  & -0.083*** \\
		0.5   & 0.9   & 10    & 0.3   & 0.55  & 0.61  & -0.063*** & 0.70  & 0.82  & -0.12*** \\
		0.5   & 0.9   & 100   & 0     & 0.74  & 0.74  & \textbf{0.001***} & 0.92  & 0.92  & \multicolumn{1}{r|}{0} \\
		0.5   & 0.9   & 100   & 0.3   & 0.73  & 0.72  & \textbf{0.01**} & 0.92  & 0.92  & -0.002*** \\
		\bottomrule
	\end{tabular}%
	\caption{Linear DGP. Median over 100 Monte Carlo simulations of the Trace of the $R^2$ between estimated and true factors. The difference is computed as: $R^2_{D^2FM}-R^2_{DFM}$. Significance level are based on a two sided Wilcoxon signed-rank test: * for $10\%$, ** for $5\%$ and *** for $1\%$.}
	\label{tab:mc_sim_l}%
\end{table}%

\begin{table}[htbp]
	\centering
	\begin{tabular}{|rrrr|rrl|rrl|}
		\toprule
		\multicolumn{4}{|c|}{\textbf{Factors}} & \multicolumn{6}{c|}{1} \\
		\midrule
		\multicolumn{4}{|c|}{\textbf{Sample}}  & \multicolumn{3}{c|}{50} & \multicolumn{3}{c|}{200} \\
		\midrule
		\multicolumn{1}{|c}{$\alpha$} & \multicolumn{1}{c}{$\rho$} & \multicolumn{1}{c}{N vars} & \multicolumn{1}{c|}{Missings} & \multicolumn{1}{c}{D$^2$FM} & \multicolumn{1}{c}{DFM} & \multicolumn{1}{c|}{Diff.} & \multicolumn{1}{c}{D$^2$FM} & \multicolumn{1}{c}{DFM} & \multicolumn{1}{c|}{Diff.} \\
		0     & 0.5   & 10    & 0     & 0.849 & 0.63  & \textbf{0.223***} & 0.88  & 0.66  & \textbf{0.217***} \\
		0     & 0.5   & 10    & 0.3   & 0.791 & 0.55  & \textbf{0.245***} & 0.827 & 0.61  & \textbf{0.22***} \\
		0     & 0.5   & 100   & 0     & 0.908 & 0.72  & \textbf{0.187***} & 0.911 & 0.76  & \textbf{0.154***} \\
		0     & 0.5   & 100   & 0.3   & 0.906 & 0.7   & \textbf{0.208***} & 0.912 & 0.74  & \textbf{0.17***} \\
		0     & 0.9   & 10    & 0     & 0.93  & 0.6   & \textbf{0.335***} & 0.945 & 0.64  & \textbf{0.302***} \\
		0     & 0.9   & 10    & 0.3   & 0.914 & 0.59  & \textbf{0.323***} & 0.94  & 0.64  & \textbf{0.299***} \\
		0     & 0.9   & 100   & 0     & 0.941 & 0.61  & \textbf{0.334***} & 0.96  & 0.65  & \textbf{0.308***} \\
		0     & 0.9   & 100   & 0.3   & 0.947 & 0.61  & \textbf{0.336***} & 0.962 & 0.66  & \textbf{0.305***} \\
		0.5   & 0.5   & 10    & 0     & 0.862 & 0.59  & \textbf{0.274***} & 0.87  & 0.63  & \textbf{0.241***} \\
		0.5   & 0.5   & 10    & 0.3   & 0.787 & 0.51  & \textbf{0.276***} & 0.802 & 0.58  & \textbf{0.222***} \\
		0.5   & 0.5   & 100   & 0     & 0.913 & 0.71  & \textbf{0.199***} & 0.912 & 0.75  & \textbf{0.161***} \\
		0.5   & 0.5   & 100   & 0.3   & 0.908 & 0.69  & \textbf{0.216***} & 0.911 & 0.74  & \textbf{0.17***} \\
		0.5   & 0.9   & 10    & 0     & 0.932 & 0.59  & \textbf{0.342***} & 0.948 & 0.64  & \textbf{0.308***} \\
		0.5   & 0.9   & 10    & 0.3   & 0.909 & 0.6   & \textbf{0.314***} & 0.934 & 0.64  & \textbf{0.295***} \\
		0.5   & 0.9   & 100   & 0     & 0.929 & 0.61  & \textbf{0.322***} & 0.962 & 0.65  & \textbf{0.31***} \\
		0.5   & 0.9   & 100   & 0.3   & 0.941 & 0.61  & \textbf{0.332***} & 0.961 & 0.66  & \textbf{0.303***} \\
		\midrule
		\multicolumn{4}{|c|}{\textbf{Factors}} & \multicolumn{6}{c|}{3} \\
		\midrule
		\multicolumn{4}{|c|}{\textbf{Sample}}  & \multicolumn{3}{c|}{50} & \multicolumn{3}{c|}{200} \\
		\midrule
		\multicolumn{1}{|c}{$\alpha$} & \multicolumn{1}{c}{$\rho$} & \multicolumn{1}{c}{N vars} & \multicolumn{1}{c|}{Missings} & \multicolumn{1}{c}{D$^2$FM} & \multicolumn{1}{c}{DFM} & \multicolumn{1}{c|}{Diff.} & \multicolumn{1}{c}{D$^2$FM} & \multicolumn{1}{c}{DFM} & \multicolumn{1}{c|}{Diff.} \\
		0     & 0.5   & 10    & 0     & 0.741 & 0.51  & \textbf{0.228***} & 0.662 & 0.44  & \textbf{0.224***} \\
		0     & 0.5   & 10    & 0.3   & 0.653 & 0.43  & \textbf{0.223***} & 0.547 & 0.34  & \textbf{0.203***} \\
		0     & 0.5   & 100   & 0     & 0.927 & 0.74  & \textbf{0.191***} & 0.948 & 0.76  & \textbf{0.184***} \\
		0     & 0.5   & 100   & 0.3   & 0.871 & 0.68  & \textbf{0.188***} & 0.924 & 0.73  & \textbf{0.193***} \\
		0     & 0.9   & 10    & 0     & 0.94  & 0.61  & \textbf{0.332***} & 0.926 & 0.65  & \textbf{0.274***} \\
		0     & 0.9   & 10    & 0.3   & 0.884 & 0.61  & \textbf{0.279***} & 0.863 & 0.64  & \textbf{0.226***} \\
		0     & 0.9   & 100   & 0     & 0.978 & 0.67  & \textbf{0.309***} & 0.987 & 0.73  & \textbf{0.253***} \\
		0     & 0.9   & 100   & 0.3   & 0.974 & 0.68  & \textbf{0.296***} & 0.984 & 0.75  & \textbf{0.232***} \\
		0.5   & 0.5   & 10    & 0     & 0.735 & 0.51  & \textbf{0.227***} & 0.638 & 0.42  & \textbf{0.222***} \\
		0.5   & 0.5   & 10    & 0.3   & 0.646 & 0.43  & \textbf{0.213***} & 0.526 & 0.34  & \textbf{0.189***} \\
		0.5   & 0.5   & 100   & 0     & 0.907 & 0.72  & \textbf{0.189***} & 0.936 & 0.75  & \textbf{0.188***} \\
		0.5   & 0.5   & 100   & 0.3   & 0.85  & 0.66  & \textbf{0.191***} & 0.91  & 0.71  & \textbf{0.197***} \\
		0.5   & 0.9   & 10    & 0     & 0.942 & 0.61  & \textbf{0.332***} & 0.922 & 0.64  & \textbf{0.278***} \\
		0.5   & 0.9   & 10    & 0.3   & 0.884 & 0.62  & \textbf{0.267***} & 0.857 & 0.64  & \textbf{0.222***} \\
		0.5   & 0.9   & 100   & 0     & 0.978 & 0.67  & \textbf{0.31***} & 0.985 & 0.73  & \textbf{0.253***} \\
		0.5   & 0.9   & 100   & 0.3   & 0.974 & 0.68  & \textbf{0.298***} & 0.982 & 0.74  & \textbf{0.238***} \\
		\bottomrule
	\end{tabular}%
	\caption{Nonlinear DGP. Median over 100 Monte Carlo simulations of the Trace of the $R^2$ between estimated and true factors. The difference is computed as: $R^2_{D^2FM}-R^2_{DFM}$. Significance level are based on a two sided Wilcoxon signed-rank test: * for $10\%$, ** for $5\%$ and *** for $1\%$.}
	\label{tab:mc_sim_nl}%
\end{table}%

An additional key advantage of the D$^2$FM are the superior performances from the point of view of its computational time. In Table \ref{tab:mc_sim_et} we compare the computational time required to estimate a DFM versus a D$^2$FM. The table shows clear computational advantages in favour of the gradient based Monte Carlo approach of the D$^2$FM as compared to the OLS EM approach of the DFM, when the dataset features many variables (starting with $150$ observable variables, in the case of our experiments).

\begin{table}[htbp]
	\centering
	\begin{tabular}{|l|cc|}
		\hline
		\multicolumn{1}{|c|}{\textbf{}} & \multicolumn{2}{c|}{\textbf{\begin{tabular}[c]{@{}c@{}}Number of \\ observations\end{tabular}}} \\ \hline
		\textbf{Number of Variables}    & \multicolumn{1}{c|}{150}                                  & 300                                 \\ \hline
		50                              & \multicolumn{1}{c|}{0.48}                                 & 0.22                                \\ \hline
		150                             & \multicolumn{1}{c|}{1.29}                                 & 1.15                                \\ \hline
		300                             & \multicolumn{1}{c|}{2.90}                                 & 2.07                                \\ \hline
	\end{tabular}
	\caption{Second order polynomial DGP with 3 factors. The table shows ratio of the DFM to the D$^2$FM elapsed times taken to build and fit the model. Each elapsed time is computed as an average over 20 runs. Both models are estimated using an Intel Core i7-8750H CPU @ 2.20GHz. We use tensorflow to estimate the D$^2$FM in eager mode as opposed to graph mode to make the performance comparable.}
	\label{tab:mc_sim_et}%
\end{table}

\section{A Deep Dynamic Factor Model for Macro Data}\label{sec:proposed_model}

To apply our model to macroeconomic data in a nowcasting and forecasting exercise, we need to introduce a version of the D$^2$FM able to track and forecast developments in economic variables in real-time. We modify the model to efficiently encode mixed-frequency data with ragged edges. In fact, economic data in real time are generally not available at the same frequency -- be it weekly, monthly or quarterly --, and missing data are a feature of real-time macroeconomic datasets, due to the non-synchronous and staggered data releases of new datapoints from statistical offices. 

As in the previous section, we specify a linear mapping between the factors and the variables (see Figure \ref{fig:model_graph}), i.e.
\begin{align}
	\boldsymbol{y}_{t} &= \boldsymbol{\Lambda} \boldsymbol{f}_{t} + \boldsymbol{\varepsilon}_{t} \ . \tag{\ref*{eq:decoding}${}^\prime$} \label{eq:linear_decoding}
\end{align}
In this form, the model can be seen as a very flexible generalisation of the approach of \cite{BAI2008304} that propose to extract factors from variables as well as their squared values and their crossproducts. 

There are a few advantages to considering this simpler D$^2$FM. First, the model maintains the same level of interpretability of a standard DFM, hence making it easy to compare the two models. Indeed, this simple architecture is motivated by the recent work of \cite{rudin2019stop} that has encouraged the design of models that are inherently interpretable, as opposed to a purely `black box' approach. Second, while interpretable, the autoencoder structure allows us to introduce deep learning methods in this framework to test its potential, towards the construction of more general models. Third, the linear decoding network and the linear state-space framework allow to update in real-time the latent states in an interpretable and statistically grounded framework, by employ a standard Kalman filter. Finally, the adoption of linear filtering techniques, in turn, allows for an easy interpretation of the model forecast revisions coming from the flow of data onto the performances of the model, as in \cite{banbura2010nowcasting}.

\subsection{Missing \& Mixed-Frequency Data}\label{sec:miss_data}

We deal with missing data in two or three steps depending on the dataset. If in the pre-training when dropping missing values we are left with a few number of observations, then we first initialise missing values with a spline method. Otherwise, this step is omitted and the pre-training is carried out only on non missing data points. Second, we iterate the parameters maximisation by replacing the missing data in the full sample with fitted values obtained by conditioning on the estimated model (both parameters and latent states). Maximisation is carried out only on non-missing points, therefore the number of observations over which the gradients are computed can differ across dimensions. Finally, in the real-time online update phase (i.e. the out-of-sample procedure), we employ the Kalman filter to update the missing data, therefore accommodating for ragged edges (\citealp{banbura2010nowcasting,banbura2014maximum,camacho2012markov}).

In dealing with mixed-frequency data there are several options (\citealp[see][for example]{marcellino2010factor, foroni2013survey, blasques2016weighted}). The most popular one, when the dataset includes monthly and quarterly variables, is the \cite{mariano2003new} approximation, in which the observed quarterly variable is modelled as a partially observed monthly series. 
By assuming that the (log-)levels of the quarterly variable, $Y_{t}^q$, at the end of the quarter are the sum of an unobservable monthly counterpart $Y_{t}^m, Y_{t-1}^m, Y_{t-2}^m$ , and defining with the growth variable $y_{t}^q$ the quarter over quarter change, we have
\begin{align*}
	\label{eqmmrestrictions}
	y_{t}^q &= Y_{t}^q -  Y_{t-3}^q = (Y_{t}^m + Y_{t-1}^m  + Y_{t-2}^m ) - (Y_{t-3}^m + Y_{t-4}^m  + Y_{t-5}^m ) 
	\\ &= 
	\Delta_3 Y_{t}^m + \Delta_3 Y_{t-1}^m + \Delta_3 Y_{t-2}^m
	\\ &=
	y_{t}^m + 2y_{t-1}^m  + 3y_{t-2}^m + 2y_{t-3}^m + y_{t-4}^m 
	,
\end{align*}
where $y_{t}^m = \Delta Y_{t}^m$ denotes the unobserved month-on-month growth rate of a quarterly variable that admits the same factor representation proposed in equation \ref{eq:decoding}. In our model, this approximation is implemented by including an additional final layer to the decoding network allowing the monthly factors to be mapped into the quarterly variables. This layer has fixed weights not subject to the optimisation.

\subsection{Model Specification and Training Details}\label{sec:model_spec}

The core of the model is provided by an asymmetric autoencoder with a nonlinear multilayer encoder and a linear single layer decoding structure. Table \ref{tab:table_hyper} provides a summary of the network design choices, and reports the choices operated for each hyperparameter of our model, a number of which are selected via cross-validation. 

\begin{table}[t!]
	\centering
	\resizebox{\textwidth}{!}{%
		\begin{tabular}{ccll}
			\toprule
			\toprule
			\multicolumn{2}{c}{\textbf{Model Components}} & \multicolumn{1}{c}{\textbf{Hyperparameter }} & \multicolumn{1}{c}{\textbf{Choice taken}} \\
			\midrule
			\midrule
			\multicolumn{1}{c}{\multirow{9}[18]{*}{\textbf{Autoencoder}}} & \multirow{6}[12]{*}{\textbf{Model Structure}} & number of hidden layers & 3 \\
			\cmidrule{3-4}          &       & number of neurons for each layer & selected via cross-validation \\
			\cmidrule{3-4}          &       & penalisation & none \\
			\cmidrule{3-4}          &       & dropout layers and rates & none \\
			\cmidrule{3-4}          &       & batch norm layers & 2 included in the encoding network \\
			\cmidrule{3-4}          &       & link function & used tanh \\
			\cmidrule{2-4}          & \multirow{3}[6]{*}{\textbf{Optimization}} & size of mini batches & 100 monthly observations \\
			\cmidrule{3-4}          &       & number of epochs & 100 for each MC iteration \\
			\cmidrule{3-4}          &       & optimisation algorithm & ADAM with default parameters \\
			\midrule
			\midrule
			\multirow{4}[8]{*}{\textbf{Dynamic Equations}} & \multirow{4}[8]{*}{\textbf{Model Structure}} & number of latent states & selected via cross-validation \\
			\cmidrule{3-4}          &       & number of lags input variables & selected via cross-validation \\
			\cmidrule{3-4}          &       & number of lags for latent common states & 2 as in \cite{banbura2014maximum} \\
			\cmidrule{3-4}          &       & number of lags for idiosyncratic states & 1 as in \cite{banbura2014maximum} \\
			\bottomrule
			\bottomrule
		\end{tabular}%
	}
	\caption{Summary of model features and choices.}
	\label{tab:table_hyper}
\end{table}%

Optimisation is carried out by using ADAM \citep[see][]{Adam} with default hyperparameters and 100 epochs, both during pre-training and training. Before starting the training, ADAM is reinitialised and then is run on batches (i.e. subsamples) with size of at least 100 monthly observations (approximately 8 years, the average duration of a business cycle). In the training phase we set again the number of epochs (runs on the full sample) to 100 for each iteration of the MCMC. These iterations are used also to update the idiosyncratic distribution.

In our empirical model, parameters are initialised in a two stage approach. First, by using a Xavier initialisation -- weights in the link functions are sampled from a Gaussian distribution with zero mean and a variance of $2/(n_{in} + n_{out}$), where $n_{in}$ is the number of input units and  $n_{out}$ is the number of output units \citep[see][]{glorot2010understanding}, and then by performing a pre-training exercise using a standard autoencoder on a full dataset where the rows that contains missing data are discarded.\footnote{In particular, in the empirical application we check that at least 50 time observations are present when applying this rule. If this is not the case, then we drop time periods that have more than 20\% missing values for the corresponding features, and we fill the rest with splines (see Section \ref{sec:miss_data}).} This pre-training procedure is needed to `warm up' the chain.\footnote{Warming up a chain in deep learning refers to the process of initialising a neural network model with weights that have been pre-trained on a related task or dataset, before fine-tuning the model on the specific dataset of interest. The goal of this process is to provide the model with a good starting point for the optimisation process, as the pre-trained weights may capture useful information that can accelerate convergence and improve performance on the new task.}

\section{Encoding the US Economy in Real Time}\label{sec:empirical_app}

We now test the performances of the model presented in the previous section in forecasting, nowcasting and backcasting using a fully real-time `big' US macro dataset, and against three benchmark models:\footnote{Backcast is the estimate of the previous quarter up to the official release date; nowcast is the estimate of the current quarter up to the official release date, and forecast is the estimate of the next quarter up to the official release date. We are able to produce backcast values because the GDP is released usually 5 weeks after the end of the reference quarter, hence we use the releases of the other variables during these 5 weeks to update the backcast figure.} (i) a univariate AR(1) statistical benchmark; and (ii) a state-of-the-art DFM with two and three latent factors, estimated via quasi maximum likelihood as proposed by \cite{giannone2008nowcasting} and generalised in \cite{banbura2014maximum} (we refer to this model as DFM-EM). The model is multitarget, i.e. it is optimised against all of the variables in the dataset and not only one of them, but our discussion of the results mainly focus on US GDP. This exercise can be seen as a validation test to check whether the model is able to correctly capture the relevant features of the data generating process, and to benchmark it against other state-of-the-art models.

\begin{figure}[t!]
	\centering
	\begin{subfigure}{.49\textwidth}
		\centering
		\includegraphics[width=\linewidth]{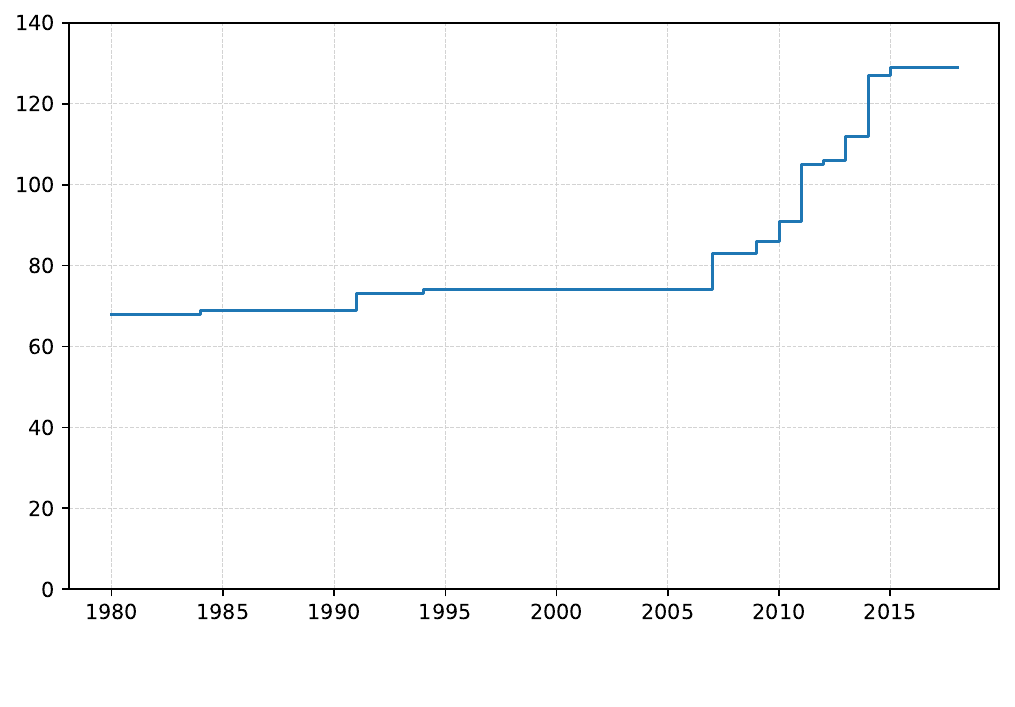}
		\caption{Number of variables.}
		\label{fig:num_variables}
	\end{subfigure}%
	\begin{subfigure}{.49\textwidth}
		\centering
		\includegraphics[width=\linewidth]{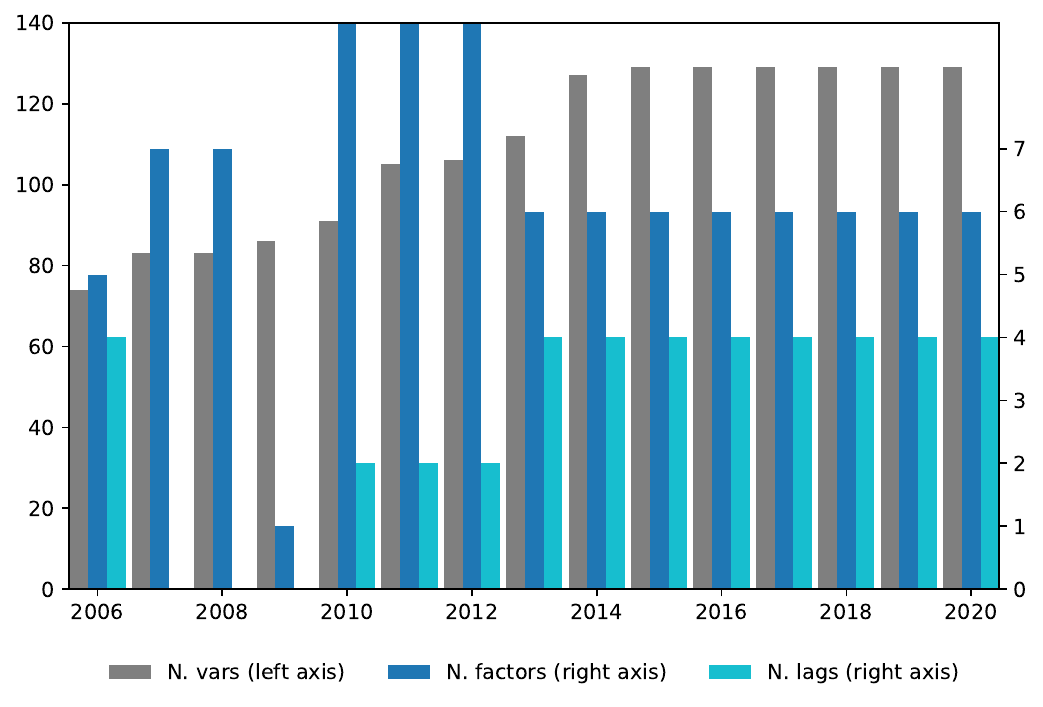}
		\caption{Number of variables, factors and lags.}
		\label{fig:cv_hyperparam}
	\end{subfigure}
	\caption{Panel (a) reports the number of variables along the entire time span taken into consideration. Panel (b) reports the number of variables, factors and lags selected via cross-validation over time. The blue line is the number of variables available for each year (left axis), red line is the optimal number of latent common states (right axis), black line is the optimal number of lags of input variables (right axis). The x-axis shows the year during which the model is used for the out-of-sample evaluation.}
	\label{fig:test}
\end{figure}

\subsection{A Real-Time `Big' Macro Dataset}

To test its capability, we estimate the model by encoding a real-time version of the full \cite{mccracken2016fred}'s FRED-MD dataset, a large macroeconomic database for the US economy, specifically designed for the empirical analysis of `big data'.\footnote{We marginally extend the dataset by including two Purchasing Managers' Indices (PMIs), since they are considered to be important indicators for nowcasting and do not get revised over time.} The cross-section of data is mixed frequency because it includes 128 monthly indicators and Real GDP, that is a quarterly variable. All the data are stationarised and standardised following the specifications in \cite{mccracken2016fred}.\footnote{In the Appendix Tables \ref{table:appendix_data1}-\ref{table:appendix_data4} provide the complete list of the variables used and their transformation code.} In Tables \ref{table:appendix_data1}-\ref{table:appendix_data4}, we report also the respective publication delay (in days) of each series. There are  substantial differences in the timeliness of different variables. Some of them are more timely (e.g. `soft' indicators or surveys), while others are released with one-two months of delay (usually `hard' data on real activity). 

The vintages in the datasets span the period from January 1980 to May 2020. Figure \ref{fig:num_variables} reports the number of variables available across time periods. We first estimate the model using the data up to December 2005, and then we perform an expanding window forecasting exercise starting form the 1$^{\text{st}}$ of January 2006. Hence our test sample goes from 1$^{\text{st}}$ of January 2006 to 31$^{\text{st}}$ of May 2020, including the Great Recession in 2007-2009. A data vintage is created every time a new time-series data point is released, and it contains all the data available up to that point in time, including also data revisions. The real-time infrastructure adapts automatically to the expanding number of variables used as input for the model. For each iteration, as new data arrive, the model is re-evaluated and outputs a sequence of backcasts-nowcasts-forecasts for GDP and all the other variables. These forecasts are conditional only to the real-time information set, i.e. only data available up to that specific point in time without taking into consideration further revisions.

Many of the hyperparameters of the model are not fixed ex-ante but are instead selected using an intensive cross-validation exercise, as reported in Table \ref{tab:table_hyper}. The real-time cross-validation exercise also provides information on the ability of the model to update its optimal hyperparameter specification over time, as new data comes in  (the validation length is set to one year). Figure \ref{fig:cv_hyperparam} shows the evolution of the number of factors and lags that are selected via cross-validation over the sample. 

\subsection{Model Evaluation}\label{sec:cv_results}

\begin{figure}[t!]
	\centering
	\includegraphics[width=1\textwidth]{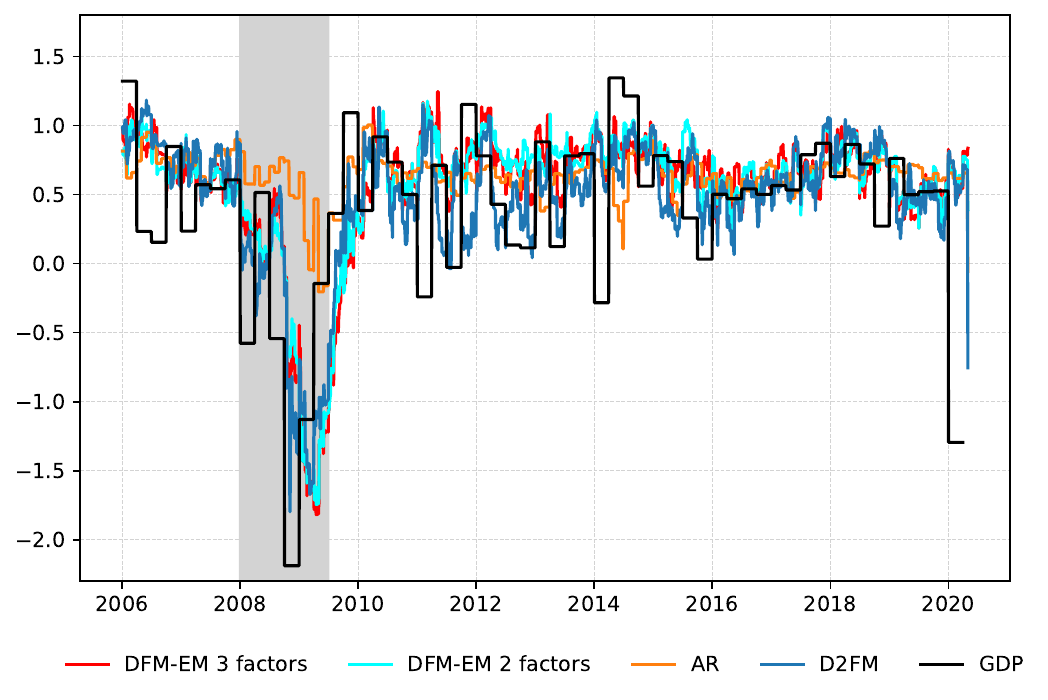}
	\caption{This Figure shows the nowcast reconstruction in real-time of the D$^2$FM, DFM-EM with 2 and 3 factors and the AR(1) versus the growth rate of the US GDP. Shaded area is the NBER recession period.}
	\label{fig:nc_rec_cv} 
\end{figure}

Figure \ref{fig:nc_rec_cv} shows the nowcast reconstruction in real-time for the D$^2$FM, the DFM-EM with 2 and 3 factors, and the AR(1) model against the realised quarterly US GDP. Overall, the D$^2$FM and the DFM-EM models provide a similar assessment of the state of the economy, in the nowcasting horizon, though the D$^2$FM is more accurate.

\begin{table}[t!]
	\centering
	\small
	\caption{Comparison of RMSEs relative to the AR(1) benchmark}
	\label{tab:RMSE}
	\begin{tabular}{l | l l | l l l | l}
		\hline \hline
		& \multicolumn{2}{c}{Forecasting} \vline & \multicolumn{3}{c}{Nowcasting} \vline & \multicolumn{1}{c}{Backcasting} \\
		\hline
		Model			&   	         30 weeks & 26 weeks & 20 weeks & 14 weeks & 8 weeks & 2 weeks \\ \hline
		D$^2$FM 			&		{\bf0.895}        &  {\bf0.906} 	   &  {\bf0.895}        & {\bf0.798}        &      {\bf0.839} & {0.832} \\
		DFM-EM 3 factors		       & 1.032         &  1.034         &  0.973     & 0.87        &      0.869  & {\bf0.826} \\
		DFM-EM 2 factors	       & 1.015	               & 1.027	   &     0.962    &   0.894       &    0.886     &    0.858 			\\
		\hline
	\end{tabular}
	\begin{tablenotes}
		\small
		\item[--] \textit{Notes}:  This table reports the RMSE of the D$^2$FM, the DFM-EM model with 2 and 3 factors relative to the RMSE of the AR(1): $RMSE(model,horizon)/RMSE(AR(1),horizon)$. Relative RMSEs are reported for different dates relative to the release date of US GDP. For example, the RMSEs at 30 weeks refers to the RMSEs 30 weeks prior to the release date.
	\end{tablenotes}
	\label{table:rmsfecv}
\end{table}

\begin{figure}[t!]  
	\centering
	\includegraphics[width=1\textwidth]{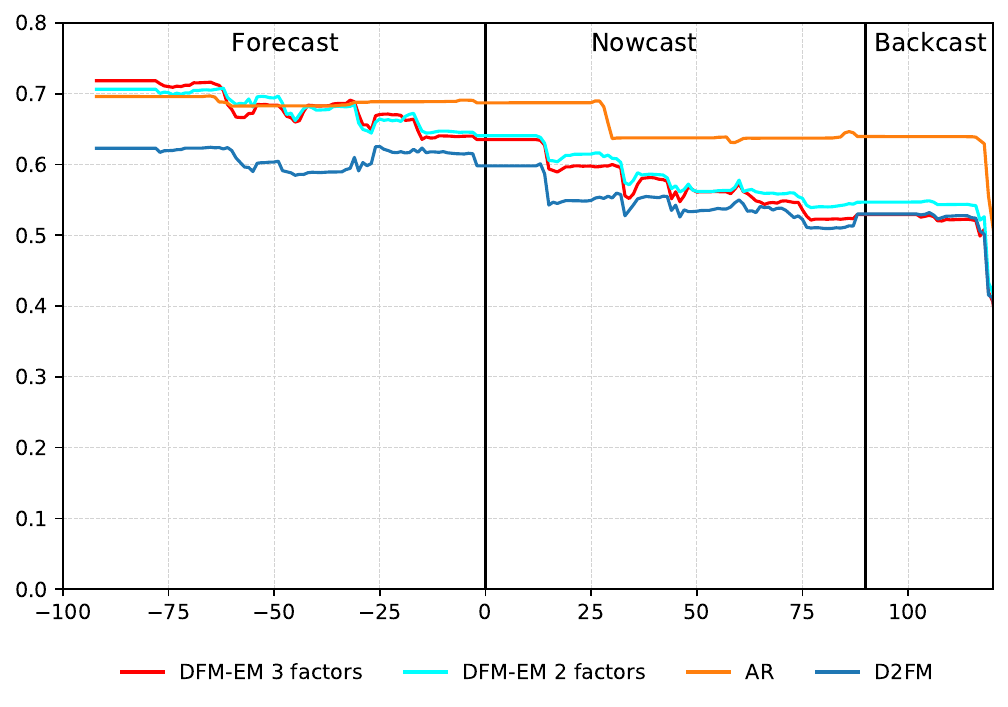}
	\caption{This Figure reports the RMSFE evolution along the quarter of the D$^2$FM model versus its competitors.}
	\label{fig:rmsfe_cv}
\end{figure}

We formally assess the performances of the model -- and of the AR(1), the DFM-EM with 2 and 3 factors -- by computing root mean square forecast error (RMSFE). This metric is updated every time the data vintage gets updated, due to a new data release. We report both an overall RMSFE (Table \ref{table:rmsfecv}) that gives us a synthetic value about the performance of each model on the entire out-of-sample set, and a dynamic RMSFE (Figure \ref{fig:rmsfe_cv}) that illustrates how the RMSFE evolves from the forecast period to the backcast period, until the day before the release.  Results indicate that the D$^2$FM is able to outperform all the competitors during the entire forecast period and for most of the nowcast period. The gain in terms of performance achieved by the D$^2$FM in these two periods is quite considerable and reflects the ability of this model to better compress the relevant information in the data, thus reducing the level of uncertainty. 

The model also delivers forecasts for all the variables in the model. Table \ref{tab:multitarget} reports the average of the RMSFEs of the D$^2$FM over all of the monthly variables, in ratio to the AR(1) RMSFEs. The D$^2$FM beats the AR(1) over all the horizons --  the backcast improves by $10\%$, the nowcast improves by $20\%$ and the forecast improves by $18\%$.\footnote{Overall, the D$^2$FM improves the prediction accuracy for roughly the $80\%$ of the monthly variables included in the dataset with respect to the AR(1).}

\begin{table}[t!]
	\centering
	\small
	\caption{Comparison of RMSEs relative to the AR(1) benchmark for monthly variables.}
	\label{tab:multitarget}
	\begin{tabular}{l | l l l | l l l | l l l}
		\hline \hline
		& \multicolumn{3}{c}{Forecasting} \vline & \multicolumn{3}{c}{Nowcasting} \vline & \multicolumn{3}{c}{Backcasting} \\
		\hline
		&	& 6 weeks      &  	   &        & 4 weeks       &    &   & 2 weeks  & \\
		\hline
		D$^2$FM		&	& {\bf 0.85}       &  	   &        & {\bf0.83}        &    &   & {\bf 0.91}  & \\
		\hline \hline
	\end{tabular}
	\begin{tablenotes}
		\small
		\item[--] \textit{Notes}:  This table reports the average RMSFE of the D$^2$FM model relative to the RMSFE of the AR(1) across all monthly variables included in the model. Relative RMSEs are reported for different dates relative to the release date of the monthly variables. For example, the RMSEs at 6 weeks refers to the RMSEs 6 weeks prior to the release date of the variable under consideration.
	\end{tablenotes}
	\label{table:multitarget}
\end{table}  

\subsection{A Real-Time Synthetic Indicator of the Business Cycle}\label{sec:AggregInd}

\begin{figure}[ht!]
	\centering
	\includegraphics[width=1\textwidth]{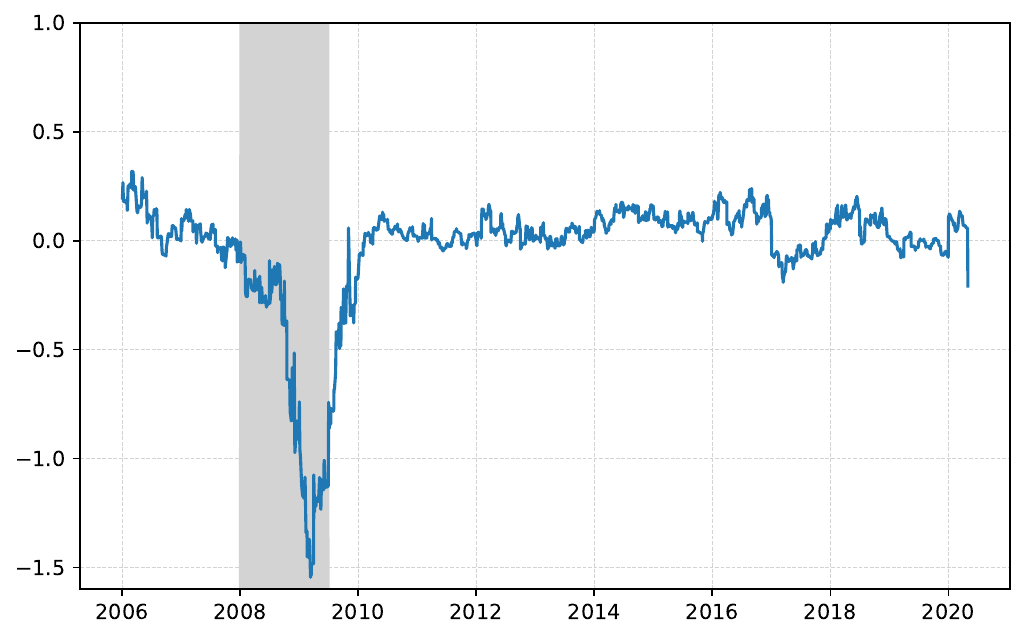}
	\caption{This Figure reports the Composite Indicator computed in real time using the D$^2$FM of section \ref{sec:cv_results}.}
	\label{fig:compositeindicator} 
\end{figure}

As a final exercise, we show how to build a composite indicator of the state of economy in real-time using the decoding map (or loadings). We do this by aggregating the latent states trough a weighting scheme. Specifically, we define the composite indicator as 
\begin{equation}
	CI = \sum_{k=1}^{r}\sum_{i=1}^{N} f_k  \frac{F_{k,i}^{2}}{||F||_{F}^{2}} \ ,
\end{equation}
where $f_k$ for $k=1,..., r$ are the common factors (the code) and $F_{k,i}$ is the matrix of the coefficient for the factor $k$ at variable $i$, as in the Equation \ref{eq:linear_decoding}, while $||F||_{F}^{2} = \sum_{k=1}^{r}\sum_{i=1}^{n} F_{k,i}^{2}$ is the squared Frobenius norm of the coefficients. The sign of the indicator is fixed to have a positive correlation with GDP. Figure \ref{fig:compositeindicator} reports the composite indicator using the real time out of sample exercise, that is shown to track well the developments in the US economy.

%
%

\section{Conclusion}\label{sec:conclusion}

The central contribution of our paper is to introduce Deep Dynamic Factor Models (D$^2$FMs) by showing how to embed autoencoders in a dynamic nonlinear factor model structure with idiosyncratic components. The equivalence between maximum likelihood estimation and minimisation of mean squared error, together with the Universal Approximation Theorem allow to conceptualise the D$^2$FMs as computationally efficient approximations of the maximum likelihood estimates of generic nonlinear factor models. 

The application of a simple version of our D$^2$FM with linear dynamic equations and a linear decoder, both in a Monte Carlo experiment, and in a big data real-time forecasting exercise shows the potential of the methodology. 

The model capability can be further generalised and employed in multiple directions. For example, one could consider empirical applications that consider a nonlinear decoding structure, or nonlinear dynamic equations for the factors. Also, the loss function could be changed to allow for quantile estimates \citep[see][for example]{koenker1978regression,chen2018quantile}. Finally, it is worth observing that the modularity and flexibility of the D$^2$FM allow to easily integrate alternative data (e.g. text data, satellite images, micro-data, etc) into the model. We leave these promising avenues of research to the future. 



\clearpage
\bibliographystyle{aer}
\bibliography{ms}
		

\newpage
\appendix
\section{Data Appendix}\label{sec:DataAppendix}

Tables \ref{table:appendix_data1}-\ref{table:appendix_data4} reports the list of variables in the dataset. 
The transformation codes (Tcode) in the Tables refers to how the variables are transformed to archive stationarity stationary. Being $X_t$  a raw series, the transformations adopted are:

\[
Z_t = 
\begin{cases}
	X_t & \mbox{if Tcode} = 1  \\
	(1-L)X_t & \mbox{if Tcode} = 2  \\
	(1-L)(1-L^{12})X_t & \mbox{if Tcode} = 3  \\
	\mbox{log}X_t & \mbox{if Tcode} = 4  \\
	(1-L)\mbox{log}X_t & \mbox{if Tcode} = 5  \\
	(1-L)(1-L^{12})\mbox{log}X_t & \mbox{if Tcode} = 6  \\
\end{cases}
\]

\begin{sidewaystable}[h!]
	\centering
	\
	\caption{Dataset (I)}
	\small
	{\begin{tabular}{| l | l | l | l | l | l | l | l | }	
			\hline \hline
			N	&	Code	&	Descriptions	& Source & Tcode & Freq & McCracken gr & Publication delay\\ \hline 
			1 & RPI & Real  Personal  Income & FRED & 5 & M & 1 & 30\\ 
			2 & W875RX1 & RPI  ex.  Transfers & FRED & 5 & M & 1 &  30\\ 
			3 & INDPRO & IP  Index & FRED & 5 & M & 1 & 14\\ 
			4 & IPFPNSS & IP:  Final  Products  and  Supplies & FRED & 5 & M & 1  & 14 \\ 
			5 & IPFINAL & IP:  Final  Products & FRED & 5 & M & 1 & 14\\ 
			6 & IPCONGD & IP:  Consumer  Goods & FRED & 5 & M & 1 &  14 \\ 
			7 & IPDCONGD & IP:  Durable  Consumer  Goods & FRED & 5 & M & 1 &  14\\ 
			8 & IPNCONGD & IP:  Nondurable  Consumer  Goods & FRED & 5 & M & 1 & 14\\ 
			9 & IPBUSEQ & IP:  Business  Equipment & FRED & 5 & M & 1 & 14\\ 
			10 & IPMAT & IP:  Materials & FRED & 5 & M & 1 & 14\\ 
			11 & IPDMAT & IP:  Durable  Materials & FRED & 5 & M & 1 & 14\\ 
			12 & IPNMAT & IP:  Nondurable  Materials & FRED & 5 & M & 1 & 14\\ 
			13 & IPMANSICS & IP:  Manufacturing & FRED & 5 & M & 1 & 14\\ 
			14 & IPB51222S & IP:  Residential  Utilities & FRED & 5 & M & 1 & 14\\ 
			15 & IPFUELS & IP:  Fuels & FRED & 5 & M & 1 & 14\\ 
			16 & CLF16OV & Civilian  Labor  Force & FRED & 5 & M & 2 & 7\\ 
			17 & CE16OV & Civilian  Employment & FRED & 5 & M & 2 & 7\\ 
			18 & UNRATE & Civilian  Unemployment  Rate & FRED & 2 & M & 2 & 7\\ 
			19 & UEMPMEAN & Average  Duration  of  Unemployment & FRED & 2 & M & 2 & 7\\ 
			20 & UEMPLT5 & Civilians  Unemployed  <5  Weeks & FRED & 5 & M & 2 & 7\\ 
			21 & UEMP5TO14 & Civilians  Unemployed  5-14  Weeks & FRED & 5 & M & 2 & 7\\ 
			22 & UEMP15OV & Civilians  Unemployed  >15  Weeks & FRED & 5 & M & 2 & 7\\ 
			23 & UEMP15T26 & Civilians  Unemployed  15-26  Weeks & FRED & 5 & M & 2 & 7\\ 
			24 & UEMP27OV & Civilians  Unemployed  >27  Weeks & FRED & 5 & M & 2 & 7\\ 
			25 & PAYEMS & All  Employees:  Total  nonfarm & FRED & 5 & M & 2 & 7\\ 
			26 & USGOOD & All  Employees:  Goods-Producing & FRED & 5 & M & 2 & 7\\ 
			27 & CES1021000001 & All  Employees:  Mining  and  Logging & FRED & 5 & M & 2 & 7\\ 
			28 & USCONS & All  Employees:  Construction & FRED & 5 & M & 2 & 7\\ 
			29 & MANEMP & All  Employees:  Manufacturing & FRED & 5 & M & 2 & 7\\ 
			30 & DMANEMP & All  Employees:  Durable  goods & FRED & 5 & M & 2 & 7\\ 
			31 & NDMANEMP & All  Employees:  Nondurable  goods & FRED & 5 & M & 2 & 7\\ 
			32 & SRVPRD & All  Employees:  Service  Industries & FRED & 5 & M & 2 &7 \\ 
			33 & USTPU & All  Employees:  TT\&U & FRED & 5 & M & 2 & 7\\ 
			\hline \hline
	\end{tabular}}
	\label{table:appendix_data1}
\end{sidewaystable}

\begin{sidewaystable}[h!]
\centering
\
\caption{Dataset (II)}
\small
{\begin{tabular}{| l | l | l | l | l | l | l | l | }
		\hline \hline
		N	&	Code	&	Descriptions	& Source & Tcode & Freq & McCracken gr & Publication delay\\ \hline 
		34 & USWTRADE & All  Employees:  Wholesale  Trade & FRED & 5 & M & 2 & 7\\ 
		35 & USTRADE & All  Employees:  Retail  Trade & FRED & 5 & M & 2 & 7\\ 			
		36 & USFIRE & All  Employees:  Financial  Activities & FRED & 5 & M & 2 & 7\\ 
		37 & USGOVT & All  Employees:  Government & FRED & 5 & M & 2 & 7\\ 
		38 & CES0600000007 & Hours:  Goods-Producing & FRED & 1 & M & 2 & 7\\ 
		39 & AWOTMAN & Overtime  Hours:  Manufacturing & FRED & 2 & M & 2 & 7\\ 
		40 & AWHMAN & Hours:  Manufacturing & FRED & 1 & M & 2 & 7\\ 
		41 & CES0600000008 & Ave.  Hourly  Earnings:  Goods & FRED & 6 & M & 2 & 7\\ 
		42 & CES2000000008 & Ave.  Hourly  Earnings:  Construction & FRED & 6 & M & 2 & 7\\ 
		43 & CES3000000008 & Ave.  Hourly  Earnings:  Manufacturing & FRED & 6 & M & 2 & 7\\ 
		44 & HOUST & Starts:  Total & FRED & 4 & M & 3 & 20\\ 
		45 & HOUSTNE & Starts:  Northeast & FRED & 4 & M & 3 & 20\\ 
		46 & HOUSTMW & Starts:  Midwest & FRED & 4 & M & 3 & 20\\ 
		47 & HOUSTS & Starts:  South & FRED & 4 & M & 3 & 20\\ 
		48 & HOUSTW & Starts:  West & FRED & 4 & M & 3 & 20\\ 
		49 & PERMIT & Permits & FRED & 4 & M & 3 & 20\\ 
		50 & PERMITNE & Permits:  Northeast & FRED & 4 & M & 3 & 20\\ 
		51 & PERMITMW & Permits:  Midwest & FRED & 4 & M & 3 & 20\\ 
		52 & PERMITS & Permits:  South & FRED & 4 & M & 3 & 20\\ 
		53 & PERMITW & Permits:  West & FRED & 4 & M & 3 & 20\\ 
		54 & DPCERA3M086SBEA & Real  PCE & FRED & 5 & M & 4 & 30\\ 
		55 & CMRMTSPL & Real  M\&T  Sales & FRED & 5 & M & 4 & 35\\ 
		56 & RETAIL & Retail  and  Food  Services  Sales & FRED & 5 & M & 4 & 30\\ 
		57 & ACOGNO & Orders:  Consumer  Goods & FRED & 5 & M & 4 & 35\\ 
		58 & ANDENO & Orders:  Nondefense  Capital  Goods & FRED & 5 & M & 4 & 35\\ 
		59 & AMDMUO & Unfilled  Orders:  Durable  Goods & FRED & 5 & M & 4 & 35\\ 
		60 & BUSINV & Total  Business  Inventories & FRED & 5 & M & 4 & 35\\ 
		61 & ISRATIO & Inventories  to  Sales  Ratio & FRED & 2 & M & 4 & 35\\ 
		62 & UMCSENT & Consumer  Sentiment  Index & FRED & 2 & M & 4 & -3\\ 
		63 & M1SL & M1  Money  Stock & FRED & 6 & M & 5 & 14\\ 
		64 & M2SL & M2  Money  Stock & FRED & 6 & M & 5 & 14\\ 
		65 & M3SL & MABMM301USM189S  in  FRED,  M3  for  the  United  States & FRED & 6 & M & 5 & 14\\ 
		66 & M2REAL & Real  M2  Money  Stock & FRED & 5 & M & 5 & 14\\ 
		\hline \hline 
\end{tabular}}
\label{table:appendix_data2}
\end{sidewaystable}

\begin{sidewaystable}[h!]
\centering
\
\caption{Dataset (III)}
\small
{\begin{tabular}{| l | l | l | l | l | l | l | l | }
	\hline \hline
	N	&	Code	&	Descriptions	& Source & Tcode & Freq & McCracken gr & Publication delay\\ \hline 
	67 & AMBSL & St.  Louis  Adjusted  Monetary  Base & FRED & 6 & M & 5 & 14\\ 
	68 & TOTRESNS & Total  Reserves & FRED & 6 & M & 5 & 36 \\ 
	69 & NONBORRES & Nonborrowed  Reserves & FRED & 0 & M & 5 & 36 \\ 
	70 & BUSLOANS & Commercial  and  Industrial  Loans & FRED & 6 & M & 5 & 20\\ 
	71 & REALLN & Real  Estate  Loans & FRED & 1 & M & 5 & 20\\ 
	72 & NONREVSL & Total  Nonrevolving  Credit & FRED & 6 & M & 5 & 14 \\ 
	73 & MZMSL & MZM  Money  Stock & FRED & 6 & M & 5 & 14\\ 
	74 & DTCOLNVHFNM & Consumer  Motor  Vehicle  Loans & FRED & 6 & M & 5 & 14\\ 
	75 & DTCTHFNM & Total  Consumer  Loans  and  Leases & FRED & 6 & M & 5 & 14\\ 
	76 & INVEST & Securities  in  Bank  Credit & FRED & 6 & M & 5 & 14\\ 
	77 & FEDFUNDS & Effective  Federal  Funds  Rate & FRED & 2 & M & 6 & -1\\ 
	78 & CP3M & 3-Month  AA  Comm.  Paper  Rate & FRED & 2 & M & 6 & 0\\ 
	79 & TB3MS & 3-Month  T-bill & FRED & 2 & M & 6 & 0\\ 
	80 & TB6MS & 6-Month  T-bill & FRED & 2 & M & 6 & 0\\ 
	81 & GS1 & 1-Year  T-bond & FRED & 2 & M & 6 & 0\\ 
	82 & GS5 & 5-Year  T-bond & FRED & 2 & M & 6 & 0\\ 
	83 & GS10 & 10-Year  T-bond & FRED & 2 & M & 6 & 0\\ 
	84 & AAA & Aaa  Corporate  Bond  Yield & FRED & 2 & M & 6 & 2 \\ 
	85 & BAA & Baa  Corporate  Bond  Yield & FRED & 2 & M & 6 & 2 \\ 
	86 & TB3SMFFM & 3  Mo.  -  FFR  spread & FRED & 1 & M & 6 & 2\\ 
	87 & TB6SMFFM & 6  Mo.  -  FFR  spread & FRED & 1 & M & 6 & 2\\ 
	88 & T1YFFM & 1  yr.  -  FFR  spread & FRED & 1 & M & 6 & 2\\ 
	89 & T5YFFM & 5  yr.  -  FFR  spread & FRED & 1 & M & 6 & 2\\ 
	90 & T10YFFM & 10  yr.  -  FFR  spread & FRED & 1 & M & 6 & 0\\ 		
	91 & AAAFFM & Aaa  -  FFR  spread & FRED & 1 & M & 6 & 0\\ 
	92 & BAAFFM & Baa  -  FFR  spread & FRED & 1 & M & 6 & 0\\ 
	93 & TWEXMMTH & Trade  Weighted  U.S.  FX  Rate & FRED & 5 & M & 6 & 2\\ 
	94 & EXSZUS & Switzerland  /  U.S.  FX  Rate & FRED & 5 & M & 6 & 2\\ 
	95 & EXJPUS & Japan  /  U.S.  FX  Rate & FRED & 5 & M & 6 & 2\\ 
	96 & EXUSUK & U.S.  /  U.K.  FX  Rate & FRED & 5 & M & 6 & 2\\ 
	97 & EXCAUS & Canada  /  U.S.  FX  Rate & FRED & 5 & M & 6 & 2\\ 
	98 & PPIFGS & PPI:  Finished  Goods & FRED & 6 & M & 7 & 16\\ 
	99 & PPIFCG & PPI:  Finished  Consumer  Goods & FRED & 6 & M & 7 & 16\\ 	
	\hline \hline 
\end{tabular}}
\label{table:appendix_data3}
\end{sidewaystable}

\begin{sidewaystable}[h!]
\centering
\
\caption{Dataset (IV)}
\small
{\begin{tabular}{| l | l | l | l | l | l | l | l | }
\hline \hline
N	&	Code	&	Descriptions	& Source & Tcode & Freq & McCracken gr & Publication delay\\ \hline
100 & PPIITM & PPI:  Intermediate  Materials & FRED & 6 & M & 7 & 16\\ 
101 & PPICRM & PPI:  Crude  Materials & FRED & 6 & M & 7 & 16\\ 
102 & oilprice & Crude  Oil  Prices:  WTI & HAVER & 6 & M & 7 & 0\\ 
103 & PPICMM & PPI:  Commodities & FRED & 6 & M & 7 & 16\\ 
104 & CPIAUCSL & CPI:  All  Items & FRED & 6 & M & 7 & 16\\ 
105 & CPIAPPSL & CPI:  Apparel & FRED & 6 & M & 7 & 16\\ 
106 & CPITRNSL & CPI:  Transportation & FRED & 6 & M & 7 &16 \\ 
107 & CPIMEDSL & CPI:  Medical  Care & FRED & 6 & M & 7 & 16\\ 
108 & CUSR0000SAC & CPI:  Commodities & FRED & 6 & M & 7 & 16\\ 
109 & CUUR0000SAD & CPI:  Durables & FRED & 6 & M & 7 & 16\\ 
110 & CUSR0000SAS & CPI:  Services & FRED & 6 & M & 7 & 16\\ 
111 & CPIULFSL & CPI:  All  Items  Less  Food & FRED & 6 & M & 7 & 16\\ 
112 & CUUR0000SA0L2 & CPI:  All  items  less  shelter & FRED & 6 & M & 7 & 16\\ 
113 & CUSR0000SA0L5 & CPI:  All  items  less  medical  care & FRED & 6 & M & 7 & 16\\ 
114 & PCEPI & PCE:  Chain-type  Price  Index & FRED & 6 & M & 7 & 30\\ 
115 & DDURRG3M086SBEA & PCE:  Durable  goods & FRED & 6 & M & 7 & 30\\ 
116 & DNDGRG3M086SBEA & PCE:  Nondurable  goods & FRED & 6 & M & 7 & 30\\ 
117 & DSERRG3M086SBEA & PCE:  Services & FRED & 6 & M & 7 & 30\\ 
118 & IPMAN & Industrial Production: Manufacturing & FRED & 1 & M & 1 & 14\\ 
119 & MCUMFN & Capacity Utilization: Manufacturing & FRED & 2 & M & 1 & 14\\ 
120 & TCU & Capacity Utilization: Total Industry & FRED & 2 & M & 1 & 14\\ 
121 & M0684AUSM343SNBR & Manufacturers' Index of New Orders of Durable Goods & FRED & 1 & M & 4 & 34\\ 
122 & M0504AUSM343SNBR & Manufacturers' Inventories, Total for United States & FRED & 1 & M & 4 & 34\\ 
123 & DGORDER & Manufacturers' New Orders: Durable Goods & FRED & 5 & M & 4 & 34\\ 
124 & CPFFM & 3-Month Commercial Paper Minus Federal Funds Rate & FRED & 1 & M & 6 & 0\\ 
125 & PCUOMFGOMFG & Producer Price Index by Industry: Total Manufacturing & FRED & 1 & M & 7 & 15\\ 
126 & ISMC & ISM Composite Index & HAVER & 1 & M & 1 & 3\\ 
127 & NAPMVDI & ISM Manufacturing: Supply Index & HAVER & 1 & M & 1 & 3\\ 
128 & SP500E & Standard \& poor 500: Price Index & HAVER & 5 & M & 8 & 0\\ 
129 & SDY5COMM & Standard \& poor 500: Dividend Yield & HAVER & 2 & M & 8 & 0\\ 
130 & SPE5COOM & Standard \& poor 500: Price/Earnings Ratio & HAVER & 5 & M & 8 & 0\\ 
131 & GDPC1 & Gross Domestic Product  & FRED & 5 & Q & 1 & 30\\ 
\hline \hline 
\end{tabular}}
\label{table:appendix_data4}
\end{sidewaystable}

\end{document}